\documentclass[manuscript,screen]{acmart}

\AtBeginDocument{%
  \providecommand\BibTeX{{%
    \normalfont B\kern-0.5em{\scshape i\kern-0.25em b}\kern-0.8em\TeX}}}

\setcopyright{rightsretained}
\copyrightyear{2022}

\usepackage{enumitem}

\newcommand\citectx[1]{\citeauthor{#1} [\citeyear{#1}]}

\begin{document}

\title{Mask-Mediator-Wrapper: A revised mediator-wrapper architecture for heterogeneous data source integration}


\author{Juraj Dončević}
\orcid{0000-0001-5221-6848}
\email{juraj.doncevic@fer.hr}
\affiliation{%
  \institution{University of Zagreb, Faculty of electrical engineering and computing, Department of applied computing}
  \streetaddress{Unska 3}
  \city{Zagreb}
  \country{Croatia}
  \postcode{10000}
}

\author{Krešimir Fertalj}
\orcid{0000-0001-7469-6168}
\email{kresimir.fertalj@fer.hr}
\affiliation{%
  \institution{University of Zagreb, Faculty of electrical engineering and computing, Department of applied computing}
  \streetaddress{Unska 3}
  \city{Zagreb}
  \country{Croatia}
  \postcode{10000}
}

\author{Mario Brčić}
\orcid{0000-0002-7564-6805}
\email{mario.brcic@fer.hr}
\affiliation{%
  \institution{University of Zagreb, Faculty of electrical engineering and computing, Department of applied computing}
  \streetaddress{Unska 3}
  \city{Zagreb}
  \country{Croatia}
  \postcode{10000}
}

\author{Agneza Krajna}
\orcid{0000-0001-8304-5463}
\email{agneza.krajna@fer.hr}
\affiliation{%
  \institution{University of Zagreb, Faculty of electrical engineering and computing, Department of applied computing}
  \streetaddress{Unska 3}
  \city{Zagreb}
  \country{Croatia}
  \postcode{10000}
}
\renewcommand{\shortauthors}{Dončević, et al.}


\begin{abstract}
This paper deals with the mediator-wrapper architecture. It is an important architectural pattern that enables a more flexible and modular architecture in opposition to monolithic architectures for data source integration systems. This paper identifies certain realistic and concrete scenarios where the mediator-wrapper architecture underperforms. These issues are addressed with the extension of the architecture via the mask component type. The mask component is detailed so it can be reasoned about without prescribing a concrete programming language or paradigm. The benefits of the new mask-mediator-wrapper architecture are analytically proven in relevant scenarios. One of the applications of the new architecture is envisioned for modern data sources integration systems backing Big data processing.
\end{abstract}



\maketitle

\section{Introduction}\label{sec1}

Research of data integration has been active for a very long time and remains ongoing to this day. It is not surprising that data integration has been examined from a multitude of angles. It started with simple ideas of monolithic multi-database systems, then evolved following the popularity of federated systems \cite{sheth_federated_1990}. The trail continued into architecture-based research with the mediator-wrapper (MW) architecture, which brought some concrete systems into existence \cite{roth_garlic_1996,roth_dont_1997,chawathe_tsimmis_1994}. At that point the idea of NoSQL systems started to appear. As data was no longer just being stored in relational databases, but also in schemaless formats in specialized DBMSs and even files, the research community started to express data integration more in terms of data sources.

Data exchange/interchange between data sources was also presented as an idea later on \cite{papakonstantinou_object_1995}. In parallel, ETL processes over multiple data sources for data warehousing also appeared and are to this day an active research topic \cite{zhang_fusion_2019, forresi_dataspace-based_2021}. 

Research has currently been exploring the idea of data lakes and how to process such large quantities of distributed, unstructured, and heterogeneous data \cite{bogatu_dataset_2020, pang_arkdb_2021}. The research landscape is also currently shifting towards graph-formatted data \cite{cappuzzo_creating_2020, da_trindade_kaskade_2020, debrouvier_model_2021, chatziantoniou_datamingler_2021} and data acquired from the Web \cite{magdy_microblogs_2020,arenas_expressive_2018, krommyda_visualization_2020}. 

It is clear that research of data integration is very unlikely to abate, especially as new kinds of potential data sources continue to be created \cite{zhou_foundationdb_2021,magdy_microblogs_2020,zimanyi_mobilitydb_2020,seidemann_chronicledb_2019, zhao_wipdb_2021, liang_cruisedb_2021}. The research is yet to produce a concrete freely usable and open-ended data source integration system. In the words of \citectx{golshan_data_2017}:

\textit{"...it is time for data integration operators to break free of end-to-end data integration systems and be available in the open source to speed up adoption and progress."}

\textit{"The first challenge […] is that progress of data integration and its application in practice are hindered by the fact that there are very few quality tools with which practitioners and researchers can freely experiment."}

Looking at the research's history in its entirety, it can be assessed that at a certain point researchers had taken a well-meaning detour to incorporate more novel systems into the field, forsaking the essentials of tooling that \citectx{golshan_data_2017} mentioned. It can be observed that the last large enterprise-wide tools for data source integration were created following the MW architecture \cite{roth_garlic_1996, chawathe_tsimmis_1994}, and that this is a good restarting point.

Of course, the scientific landscape has greatly changed since the 1990s. Today there is also a shift of discussion about the way in which views are created, or how data sources are represented. With reintroducing the MW architecture great care is taken to conform it to modern and future requirements, making it the conduit for future scientific work regarding data.

This paper is focused on the MW architecture for data source integration systems and some of its deficiencies. These deficiencies will be presented through the translation of different types of schemas in the system and how the allocation of these schemas affects the responsibilities of certain types of system components. This will show that some components have additional responsibilities, intruding on a flexible and scalable architectural design and making the system difficult to maintain and administrate. This paper then offers a solution to this quandary and presents the need for an additional component in the MW architecture, which will allow for a flexible and scalable system componentization and easier real-world use. Consequently, this paper introduces the mask component as this additional component to the existing component types. This addition prompts us to introduce an extended architectural pattern based on the MW architecture – the mask-mediator-wrapper (MMW) architecture.

\section{The mediator-wrapper architecture}\label{sec2}

The MW architecture was first envisioned as an information system architecture \cite{wiederhold_mediators_1992}, allowing a modular architecture for subtasking when numerous data sources are imposed, in opposition to monolithic architectures. This was specifically intended for information and knowledge management systems for informed decision making. 

Expanding on the idea of what is achievable using the MW architecture, \citectx{papakonstantinou_object_1995} observed its usage for exchange of data across heterogeneous information sources. \citectx{roth_dont_1997} also observed the mediator-wrapper architecture to uniformly access legacy stores through the GARLIC system \cite{roth_garlic_1996}. Similarly, the mediator-wrapper architecture was used as a basis for the TSIMMIS project \cite{chawathe_tsimmis_1994}. \citectx{garcia-molina_database_2008} put the mediator-wrapper architecture firmly into the context of data source integration systems.

The MW architecture in the most general sense is an architectural pattern, consisting of mediator and wrapper components, used to query and acquire data from multiple data sources.

The wrapper component is directly connected to a data source and acts as a standardized interface to that data source. The wrapper wraps (or encapsulates) the data source for further use throughout the rest of the system, effectively making it the only component in direct contact with the data source. To ensure such functionalities, the wrapper must be able to translate queries, data and metadata coming to and from the data source, as well as the layers above.

The mediator component is architecturally situated above the wrappers. The mediator’s task is to connect multiple wrappers and integrate their data and metadata. Because data, metadata and queries are logically intertwined, the mediator also must have the ability decompose and allocate queries to its connected wrappers.

\begin{figure}[h]
\centering
\includegraphics[width=0.5\textwidth]{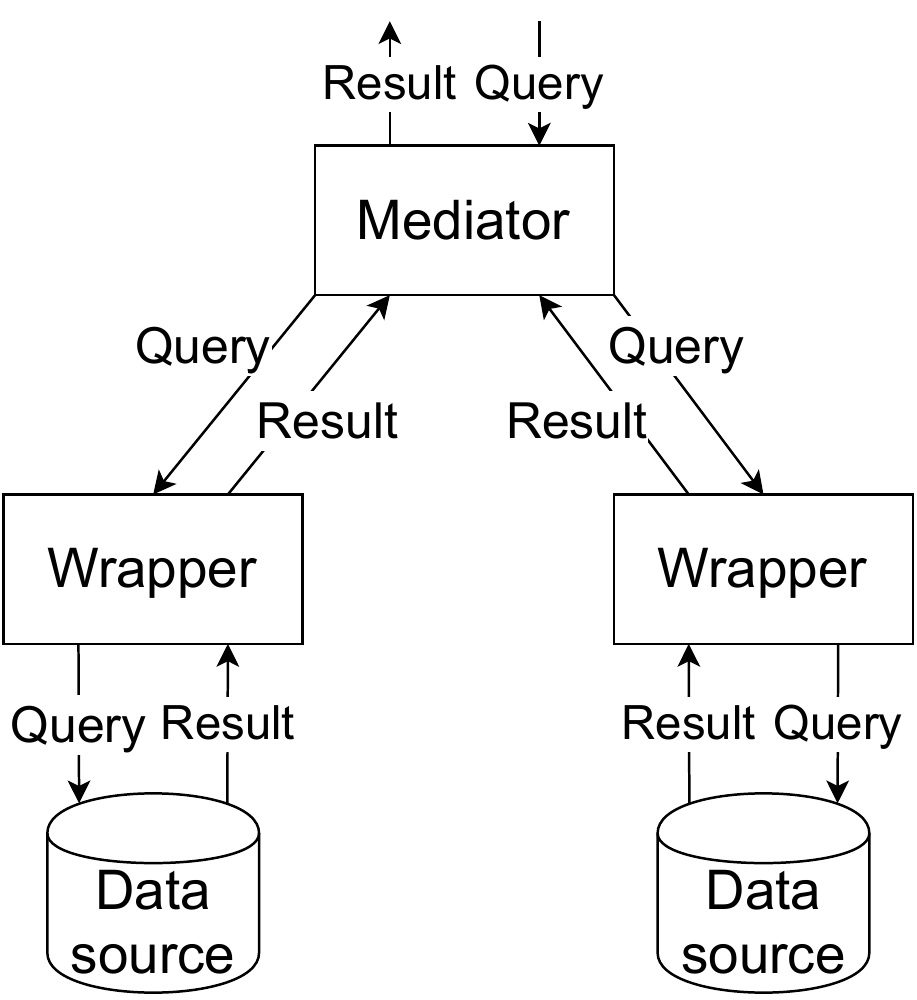}
\caption{Relationship of mediator and wrapper components \cite{garcia-molina_database_2008}}\label{fig1}
\end{figure}

Certain aspects of the MW architecture can be clarified by following the top-to-bottom flow of data as shown in Fig.~\ref{fig1}. The mediator receives a query which is then propagated accordingly to its connected wrappers. Not all wrappers need be included in a query, as all the data required by the query might not be in all the data sources. The queried wrappers then translate the queries according to their data source’s schema and querying language. The returned result is then translated back into the system standardized result format and propagated to the mediator and above.

\begin{figure}[h]
\centering
\includegraphics[width=0.9\textwidth]{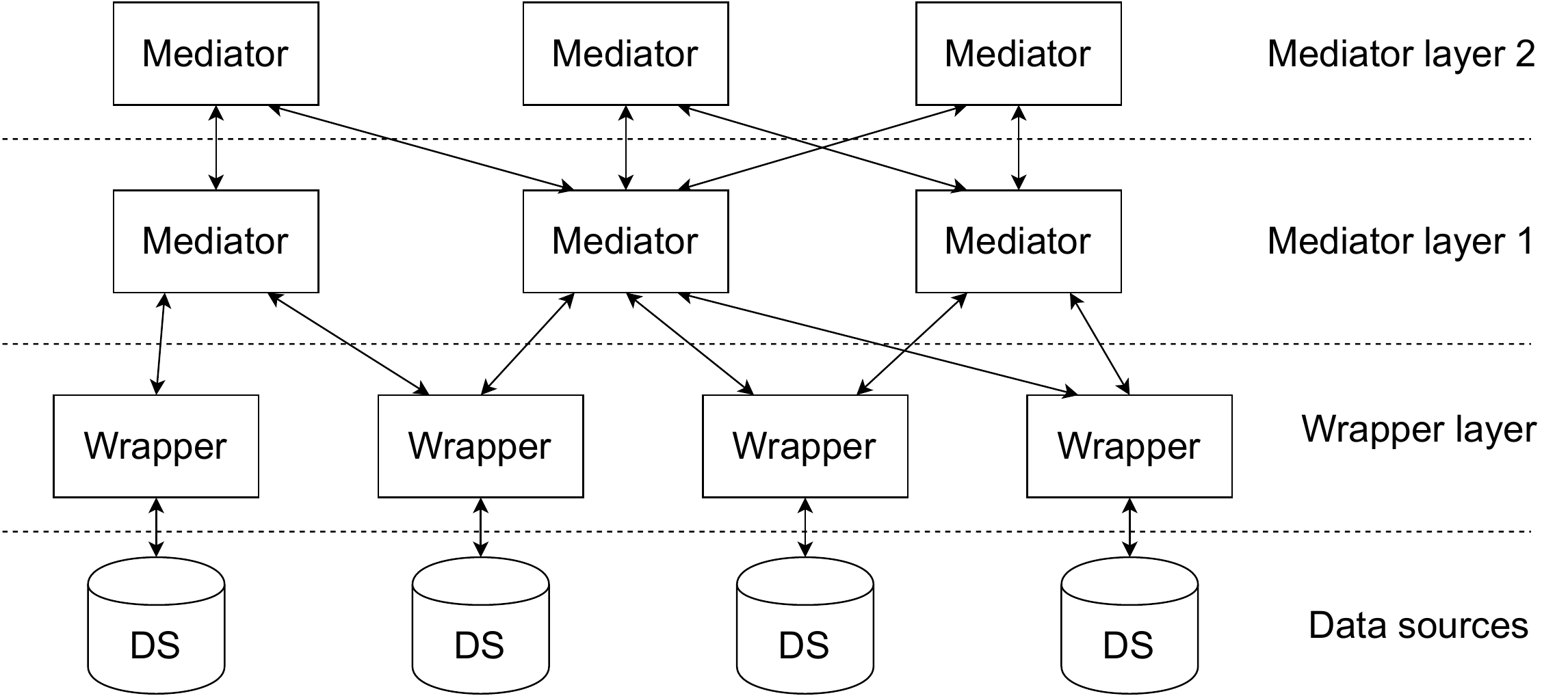}
\caption{MW architecture with layered mediators \cite{ozsu_principles_2011} }\label{fig2}
\end{figure}

This pattern of interaction is the basis for the complete MW architecture. A more global view is shown in Fig.~\ref{fig2}. The data sources to be integrated are at the lowest layer. Each data source is covered by a single wrapper via a direct connection. As an example, a system where each wrapper operates over a single data source is displayed (Fig.~\ref{fig2}), although \citectx{ozsu_principles_2011} display a possibility of a wrapper operating over multiple data sources. It can be observed that the “one wrapper – one data source” setting gives more agility for appending new data sources to the integration system, as it allocates the responsibility of an overseeing data sources to each wrapper separately and thus balances the workload. It is also interesting to comment that this component setting is better suited for systems being built bottom-up \cite{busse_federated_1999, ozsu_principles_2011}, where data sources are expected to be appended and the global data overview to change. 

The first layer of mediators is located directly above the wrapper layer. Fig.~\ref{fig2} displays their relationship in a form where each mediator in this layer can be connected to multiple wrappers and multiple wrappers can be connected to a single mediator. This is in line with the MW architecture displayed in \citectx{ozsu_principles_2011}. \citectx{papakonstantinou_object_1995} and \citectx{jurczyk_dobjects_2012} displayed an architecture in which each mediator of the first mediator layer is connected to just one wrapper and vice-versa; showing that this is also a feasible solution in cases where mediators are only needed for translation. The first mediator layer can be used to mediate between wrappers over paradigmatically similar data sources or data sources that have an overlapping or connected domain.

The second and upper layers of mediators can be used to raise the level of abstraction. The mediators of the upper layers are used to mediate between mediators of the lower layers, thus possibly encompassing multiple different data sources. Such a layering strategy is used by \citectx{moura_integrating_2005} in a form of special and central mediators to organize and distribute processing load, which can also be a beneficial effect if components are run on different machines. On the other hand, \citectx{chawathe_tsimmis_1994} uses layering to enable localized logical management of data sources. This layering strategy is also proposed by \citectx{ozsu_principles_2011}.

There are also proposed and implemented systems with one monolithic mediator \cite{roth_dont_1997, hongzhi_wang_effective_2003, chang_applying_2011, shao_integrated_2013, garg_integration_2015, schmatz_interface_2018}. This can be found to be an ample, quick, and expedient solution if the number of connected data sources is not large or expected to rise. If the number of connected data sources rises, then the processing load on that single mediator is increased and this can easily lead to increased latency when querying any of the connected data sources through the integration system. These types of systems can be covariantly classified as multidatabase integration systems \cite{doncevic_database_2020}, taking note that the data source component translation is distributed (assigned to the wrappers). It should also be noted that in these cases the mediator component is not really a component, but rather a software module.

Recently, \citectx{sethi_presto_2019} have also focused on creating and maintaining a concern-oriented architecture system. Their \textit{workers} each have a \textit{Data Source API} in similarity with wrappers. The mediator's functionalities are assigned to a single \textit{coordinator} acting as the query entry-point and planner, and another worker to join query results (akin to a reducer node in MapReduce). 

\subsection{On the roles of mediator-wrapper components}\label{sec2-1}

Considering the previously mentioned roles and interactions of wrappers and mediators in the MW architecture, it can be determined what kind of properties these components should have, or to which rules they should adhere. In a general sense any system component should satisfy the following conditions \cite{meyer_grand_2003}:
\begin{itemize}
    \item It can be used by other software elements, its “clients”. 
    \item It possesses an official usage description, which is sufficient for a client author to use it.
    \item It is not tied to any fixed set of clients.
\end{itemize}

This set of conditions can be expanded to determine what conditions should a wrapper or mediator specifically meet. For wrappers in a data source integration system, by following the example of \citectx{roth_garlic_1996} and \citectx{roth_dont_1997} (in their case the GARLIC system), the following rules (goals) are set:

\begin{enumerate}[label=\textbf{RW\arabic*}, leftmargin=*]
    \item \label{rw1} The start-up cost to write a specific wrapper should be small. The wrapper itself can be constructed quickly with little need for prior knowledge of the data source integration system internal structure. There is a basic service upon which a specific wrapper is built upon.
    \item \label{rw2} Wrappers should be able to evolve. Incremental upgrades to the wrapper should be possible.
    \item \label{rw3} Wrappers should be modular and independent. Wrappers for new data sources can be integrated into the existing data source integration system without disturbing user applications, and other wrappers or components.
    \item \label{rw4} Wrappers should be participants in query planning. The wrapper may use whatever knowledge it has about a repository’s query and specialized search facilities to dynamically determine how much of a query the repository is capable of handling.
\end{enumerate}

For mediators in a data source integration system, following the ideas of \citectx{wiederhold_mediators_1992}, the following rules are set:
\begin{enumerate}[label=\textbf{RMe\arabic*}, leftmargin=*]
    \item \label{rme1} Structuring mediators into hierarchies should not lead to problems.
    \item \label{rme2} Mediators should drive transformations. Mediators are there to accommodate the need for data and metadata restructuring. Queries are also affected by this restructuring.
\end{enumerate}

\subsection{On schema hierarchies in the mediator-wrapper architecture}

\begin{figure}[h]
\centering
\includegraphics[width=0.9\textwidth]{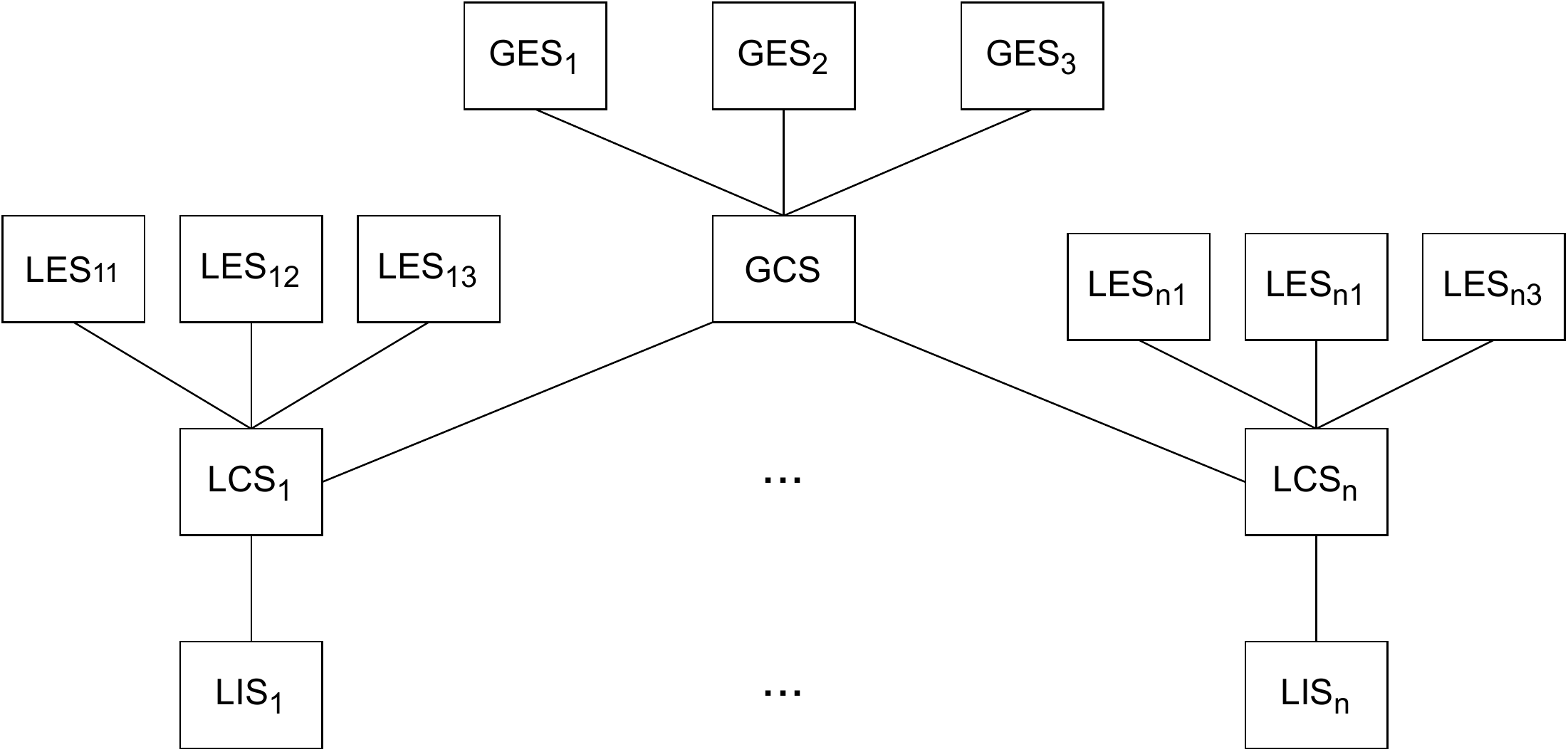}
\caption{A schema hierarchy \cite{ozsu_principles_2011} }\label{fig3}
\end{figure}

One of the advantages of using a MW architecture is the ability to modularly translate schemas by using the architecture’s components themselves. To better understand these specifics, a generic example of a schema type hierarchy is displayed in Fig.~\ref{fig3}, which shows all the possible schema types and their possible relationships. This is in multiple forms explained by \citectx{ozsu_principles_2011}.

Starting bottom-up, the first type of schema is a local internal schema (LIS). The LIS is the schema found in the connected data source itself, defined in the data source’s native form. For the data source integration system to be able to work on the connected data source it must translate the LIS to a more generic and adaptable form that is used system-wide – this is the local conceptual schema (LCS). The LCS can then be translated into a local exported schema (LES). The LES is for all intents and purposes a partial or transformed schematic view of the LCS. As the data source integration system does not use the LES for its internal functioning, the LES can be described in an entirely different form and presented to the user. The global conceptual schema (GCS) is created by integrating the local conceptual schemas. In turn, the GCS can also be exported to the user in multiple forms, just like the LES. Such an exported schema is called a global exported schema (GES).

In a multidatabase integration system these schemas are all found in the same integration component in the form of metadata and are generated by modules. On the other hand, in a MW architecture data source integration system these schemas are worked on gradually. This is done through the system’s wrapper and mediator layers; each layer creating a more encompassing global schema or creating new forms of exported schemas.

\begin{figure}[h]
\centering
\includegraphics[width=0.7\textwidth]{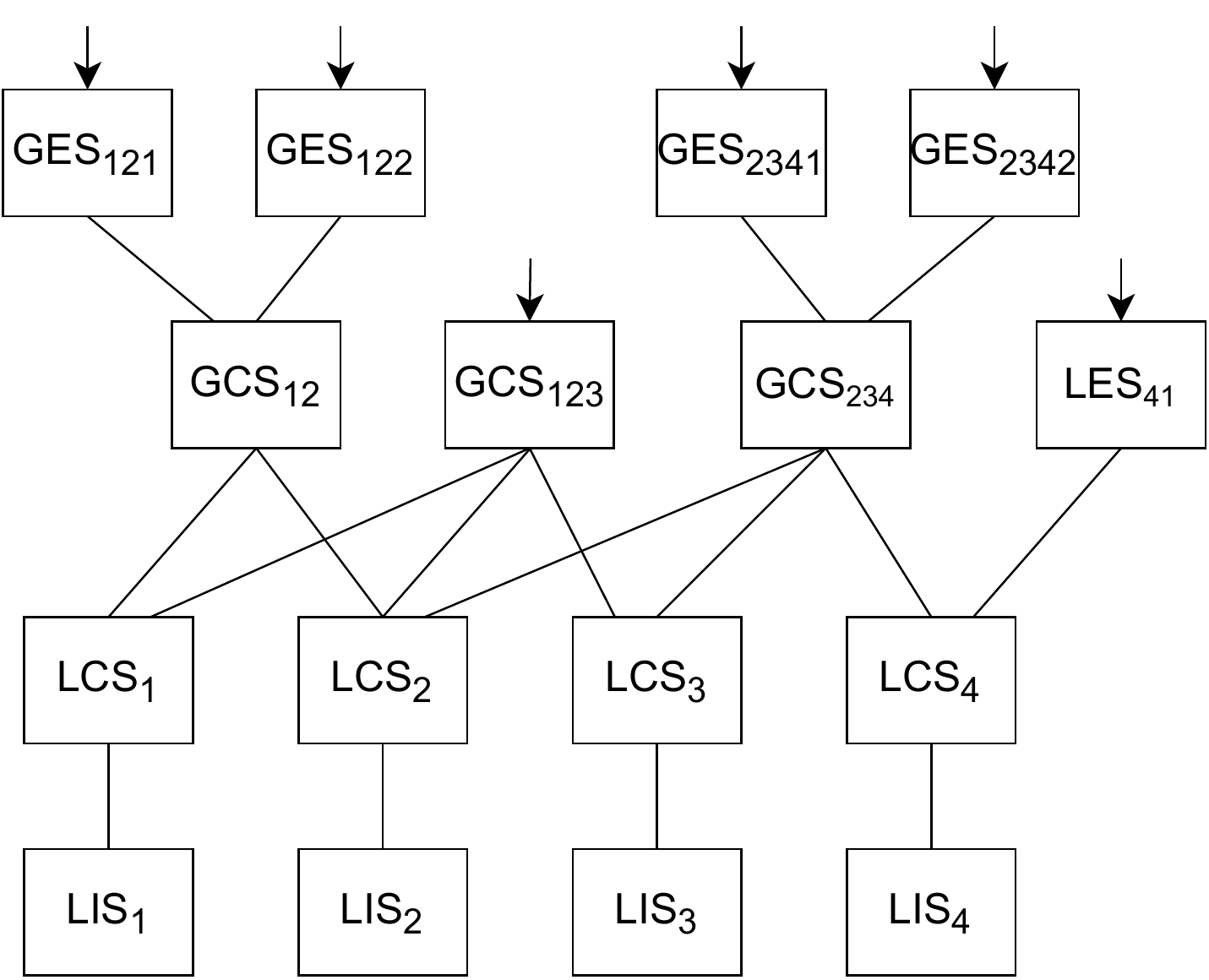}
\caption{An example of a system-wide schema hierarchy}\label{fig4}
\end{figure}

As an example, a certain system-wide schema hierarchy is presented in Fig.~\ref{fig4}; with the schemas’ relationships in accordance with the former explanation. The arrows on top of the schemas in Fig.~\ref{fig4} demonstrate which of them can be accessed by a user.

\begin{figure}[h]
\centering
\includegraphics[width=\textwidth]{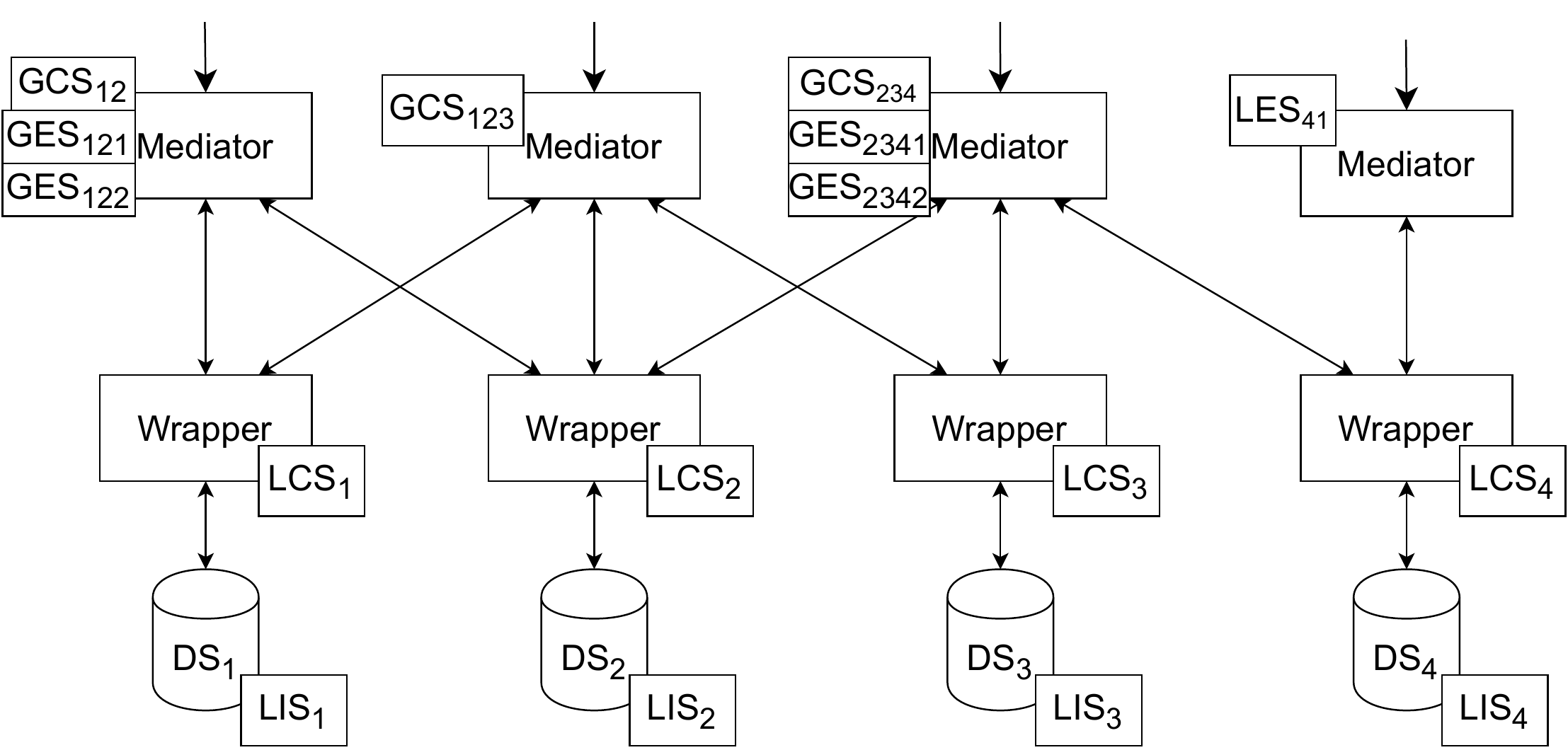}
\caption{An exemplified assignment of schemas to a MW system with a single mediator layer }\label{fig5}
\end{figure}

To set a concrete example this paper displays how schemas should be assigned to MW components. Setting this example for the continuation of this paper, the schemas from the exemplified hierarchy (Fig.~\ref{fig4}) can now be assigned to MW architecture components. The first possible assignment of schemas is to a system where there is a single mediator layer, as displayed in Fig.~\ref{fig5}. The LISs are positioned in the connected data sources. The wrappers then form their individual LCSs based on their connected data source’s LIS. The wrappers’ LCSs are then used by the mediators to create their GCSs. In this example an aforementioned case of a mediator connected to a single wrapper is also displayed. This mediator generates a LES, thus this mediator is only used for translation. The other mediators, along with their GCSs, generate GESs. A mediator can be used to create GESs to remove the need for another architectural layer of mediators above. Of course, this might decrease system latency, but will increase the complexity of mediator components as they now must manage multiple user role access. 

\begin{figure}[h]
\centering
\includegraphics[width=\textwidth]{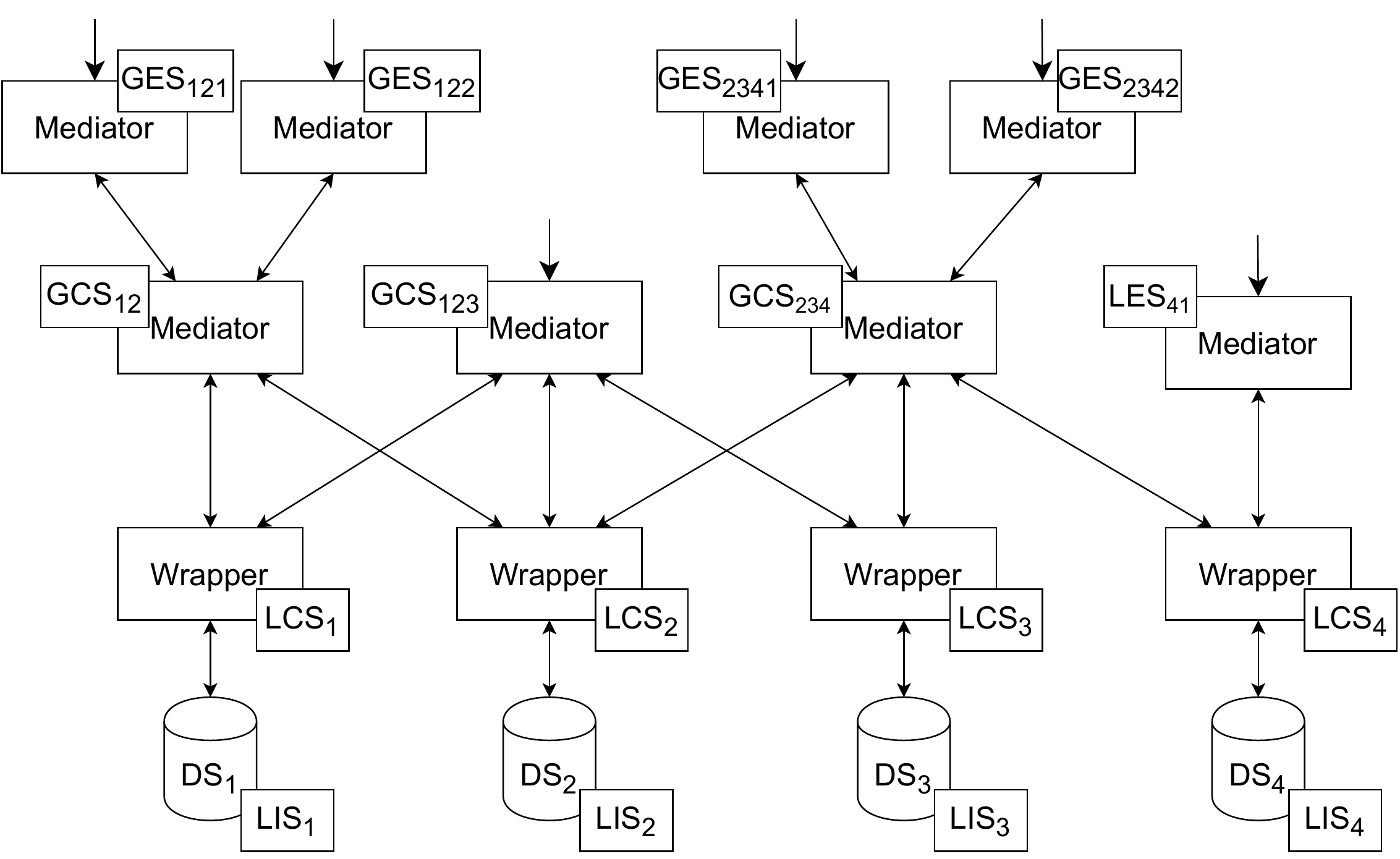}
\caption{An exemplified assignment of schemas to a MW system with an exporting mediator layer}\label{fig6}
\end{figure}

Another example of a schemas’ assignment is displayed in Fig.~\ref{fig6}. In this example there is another mediator layer on top of the architectural hierarchy. These mediators are used exclusively for exporting schemas; similar to the translating mediator. In this alternative each mediator exports just one form of GES, thus reducing their task-base and reducing the required complexity for multiple user role management. In other words, each mediator could have just one form of a data-accessing user.

\section{Problems with the mediator-wrapper architecture}

So far, mentioning the way in which users connect and use these systems has been omitted. Multiple authors have opted for connecting users to the system via specialized applications which are system-specific \cite{moura_integrating_2005, chang_applying_2011, shao_integrated_2013}. Such applications connect directly to the highest mediator layer (or the integration layer in case of an alternative architecture).

This puts additional responsibilities on the mediators in the higher layers. These mediators not only have to mediate schemas from lower layers, but also manage their GESs, as shown in the example of Fig.~\ref{fig5}. Opposed to this, the exporting mediators shown in Fig.~\ref{fig6} seem like a better solution due to the functional responsibility being shared among multiple mediator components. This also has its issues. RMe1 prohibits the mediator from exporting data in a format that is not internally used by the system itself. Meaning that data translation is going to have to be done in a user application. This breaches the system’s separation of concerns, leading to client applications having to do the translations. It seems that this responsibility cannot be shared among components of the mediator type. Therefore, we add a third rule for mediators:

\begin{enumerate}[label=\textbf{RMe\arabic*}, leftmargin=*]
    \item[\textbf{RMe3}] \label{rme3} Mediators should be used to mediate, not to represent.
\end{enumerate}

This problem is further exacerbated when one takes notice of the user applications usually implemented alongside these systems. Although user applications generally display just one format of data, it is interesting to notice the variety of data formats that have been used as presentational in different systems – from JSON collections \cite{atzeni_uniform_2014,vathy-fogarassy_uniform_2017}, XML documents \cite{li_xml-based_2005}, to tabular data \cite{kozankiewicz_integration_2004, lawrence_integration_2014}. The way of access can also be varying – a JDBC API \cite{lawrence_integration_2014}, web applications built onto the system \cite{vathy-fogarassy_uniform_2017} and even a web API \cite{atzeni_uniform_2014}. 

This is also the case with state-of-the-art databases and frameworks designed with specific representations of data sources in mind. Some authors still show a preference for an SQL interface \cite{sethi_presto_2019, zimanyi_mobilitydb_2020, abuzaid_diff_2021, seidemann_chronicledb_2019}, while others prefer a key-value \cite{liang_cruisedb_2021,zhao_wipdb_2021,zhou_foundationdb_2021}, graph \cite{da_trindade_kaskade_2020, chatziantoniou_datamingler_2021,debrouvier_model_2021}, semantic web \cite{arenas_expressive_2018, krommyda_visualization_2020}, an XML \cite{li_flashschema_2020}, flattened data \cite{lam_automated_2021}, or a plain-text interface \cite{gkini_-depth_2021}. \citectx{pang_arkdb_2021} have also shown a system with three types of data representations: an object storage (REST API), file storage, and NoSQL tablestore service. \citectx{benedikt_balancing_2021, qin_making_2020} have also shown that data representation (views) is becoming a key factor in data handling.

With the increase of data format variety, it is becoming more apparent that a data source integration system, as a singular data source, will itself have to support data representation in different formats. It is important to note that data, schemas, and queries face this same issue equally.

A more general point is that the MW architecture in its current state diverges from the idea of a clean system architecture. The clean architecture principles of architectural layering, separation of concerns, managing dependencies, control flow and testability are a solution to achieve a flexible and largely scalable system \cite{martin_clean_2017, ivanics_introduction_2017}. Such a system is an expected requirement for gathering and managing large amounts of data from multiple sources.

\section{Extending the mediator-wrapper architecture}

It is evident that currently in the MW architectural pattern the responsibility of representing data, schemas, and queries cannot be assigned to any of the existing component types without assigning too much responsibility onto them. For this reason, the system designer is forced to decide whether to assign this responsibility to the highest mediator layer or a user application.

Due to the nature of the problem being the assignment of a system functionality to a component type, and all existing component types being finely utilized via their given rules, it has become obvious that there is a componential gap in the upper layer of the MW architecture. In other words, due to RMe3 there is a task that no component type is adequate to additionally handle. Hence, there is a requirement for another type of system component that could take on the responsibility of representing system data.

\begin{figure}[h]
\centering
\includegraphics[width=\textwidth]{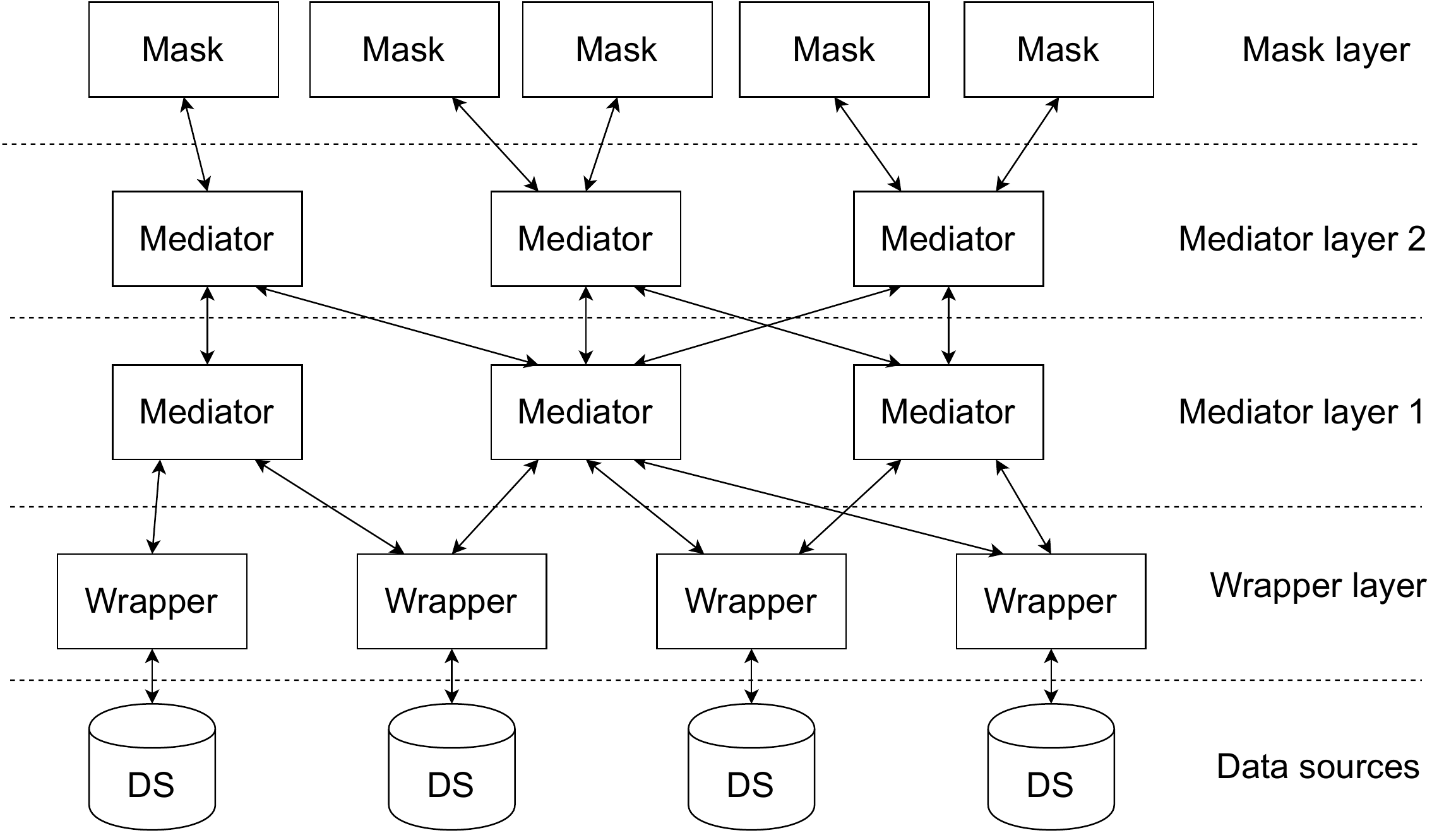}
\caption{The MMW architecture with layered mediators}\label{fig7}
\end{figure}

Therefore, in this article we propose a new theoretical component, which we name a mask. A mask masks the system at a certain point in the schema hierarchy into a representational form that can be easily handled by users, effectively taking on the responsibility of representing the system. The mask should be placed at the top of the architectural hierarchy; positioned between the users and highest mediator layer. Placing the masks on top of the architectural hierarchy effectively creates a mask layer. Consequently, this extended variant of the MW architecture is called a MMW architecture (Fig.~\ref{fig7}).

Using the mask, the system’s representational logic is decoupled from the system’s mediation logic and the user’s application logic. Furthermore, by adding the mask as a system component type the system has a finer separation of responsibility and gains benefits that help expand and simplify its usability. If a mask supporting a form of standardized technological access to data is implemented, then access to the system becomes available to a wide variety of applications implemented over that standard of access. 

Observing the mask with an implementation example, one could implement a mask in the form of a web representational state transfer (REST) service with requests over uniform resource locators (URL) returning resources in JavaScript Object Notation (JSON), akin to the system access shown in \cite{atzeni_uniform_2014}. In this way any application built to send requests to a REST service and receive its responses can now be used as a client application.

\begin{figure}[h]
\centering
\includegraphics[width=0.4\textwidth]{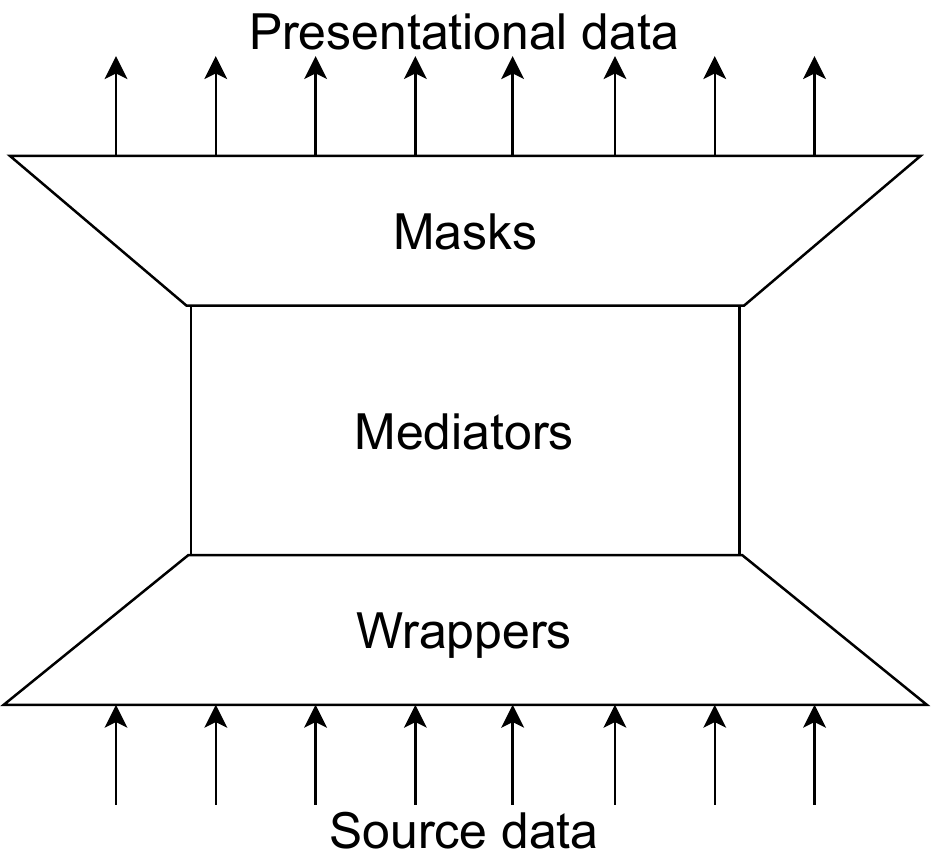}
\caption{Stylistic view of the MMW architecture}\label{fig8}
\end{figure}

Another interesting way to look at a mask component is to imagine it as an inversed wrapper Fig.~\ref{fig8}. While the wrappers concern themselves with adapting the source data from the outside world to accommodate the data source integration system’s standard, the masks concern themselves with adapting the standardized data to accommodate the outside world. Also, wrappers import data from multiple sources, while masks export data to multiple destinations. Hence, the data source integration system can now be seen as a single logical point of collecting, transforming and providing data in various formats.

As was the case with wrappers and mediators, the rules for masks are set as:
\begin{enumerate}[label=\textbf{RMa\arabic*}, leftmargin=*]
    \item \label{rma1} A mask should be positioned at the top of the architecture.
    \item \label{rma2} A mask only connects to a single mediator.
    \item \label{rma3} A mask is used for representational purposes – representing a schema, querying data, representing result data.
\end{enumerate}

\ref{rma1} follows from the consensus that the presentation layer in system architectures is positioned at the top (furthest on the user-side). The mask, its use being representation, is the system’s presentation layer. 

\ref{rma2} follows from the reversal of its statement. If the mask could connect to multiple mediators, then it would need to also apply mediation – breaking the separation of responsibilities among the components. Hence, a mask is allowed to connect to just one mediator, and all the mediation is left to the mediators.

\ref{rma3} states a set of basic functional requirements that are expected of most data access systems. This rule articulates that the mask component doesn’t in any way diminish the system’s functionality. 

\subsection{The mask’s effect on the system schema hierarchy}

\begin{figure}[h]
\centering
\includegraphics[width=\textwidth]{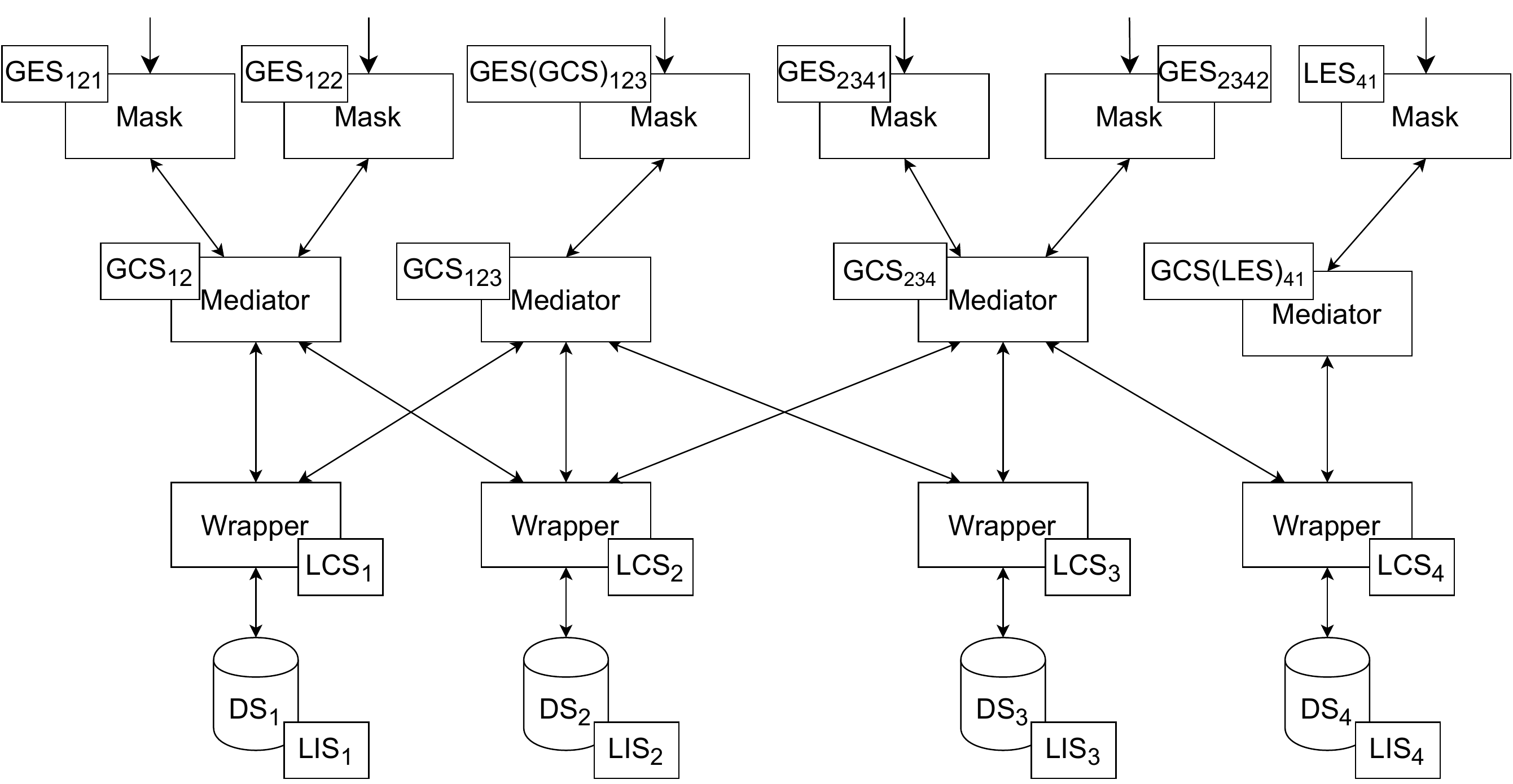}
\caption{An exemplified assignment of schemas to a MMW system}\label{fig9}
\end{figure}

To show that the addition of masks affects only the mediators in the higher layers and decreases these mediators’ responsibilities, in Fig.~\ref{fig9} the assignment of schemas from the system-wide schema hierarchy from Fig.~\ref{fig4} is shown.

The wrappers themselves and their schemas have remained unchanged, but there is a significant difference above the first mediator layer. It is important to note that the placement of prior existing components has not been changed – all the mediators still connect to the same wrappers and the mediators all operate over the same GCSs. Analogous to the examples shown in Fig.~\ref{fig5} and Fig.~\ref{fig6}, the mediator components of the (now only existing) mediator layer operate over their respective GCSs. The mask components have been assigned all the GESs.

There is a noteworthy schema rename in the example of Fig.~\ref{fig9}, for what was originally GCS\textsubscript{123}. As the GCS\textsubscript{123} itself was an exported schema in prior examples, in this example the schema might be in a fundamentally different format. Hence, the schema operated over in the mask cannot be named the same as the schema in the mediator. To mark this change, what was once GCS\textsubscript{123} used for exporting is now GES\textsubscript{123} – a fully-fledged exported schema.

A similar effect can be seen in the case of the mediator that in Fig.~\ref{fig5} and Fig.~\ref{fig6} operated over the LES\textsubscript{41}. As this mediator’s schema is not directly exported, it has been renamed as GCS\textsubscript{41}, although it currently only incorporates LCS\textsubscript{4}. The mask component above this mediator has taken the responsibility of representation (exporting) and has consequently been assigned LES\textsubscript{41}.

There is also an interesting case in Fig.~\ref{fig9} concerning the translation of schema LCS\textsubscript{4} to the upper layers (via LCS\textsubscript{4}, GCS\textsubscript{41}, LES\textsubscript{41}) and the components used for this task. The previously exporting mediator of LES\textsubscript{4} from Fig.~\ref{fig5} and Fig.~\ref{fig6} has been preserved. This mediator’s schema has also been renamed to GCS\textsubscript{41}, as stated earlier. For the moment ignoring the \ref{rma2}, it can be questioned whether this purely translational mediator is even required. Rightly so, if it is additionally considered that the system probably uses standardized interfaces for inter-component communication. It could be concluded that the mask with LES\textsubscript{41} could be connected directly to the wrapper with LCS4.

But this is not the case, as the mediator with GCS\textsubscript{4} (formerly LES\textsubscript{4}) must be preserved. The reason for this statement is two-fold from the angle of system design. Firstly, the mediator is not only used for translation, but also enables transformations within the schema itself (as is stated by \ref{rme2}). Connecting the mask directly to the wrapper, although feasible, would disable the system to apply further transformations on schema LCS\textsubscript{4}. Secondly, the benefit of using a MW architecture, and by extension our own, is the ability to append data sources after the system has been set up. Connecting the mask directly to the wrapper leaves the system without a mediator to mediate between the wrapper with LCS\textsubscript{4} and any additional to-be-connected wrapper. Because of this the system would lose the beneficial property of being (completely) appendable.

This is an example of how the \ref{rma2} preserves not only the component hierarchy of the architecture, but also the properties of the system itself. 

\subsection{On the implementation of a mask}

The general practice in this paper up to this point was to analyze the mask as a generic black box component and explain how it would work in synthesis with other system components. To expand the idea of the mask even further it can no longer be observed just as a black box. The possible inner workings of a mask give the ability to distill this architectural component even further in terms of design and development. As with most software systems, the mask, a miniature system itself, can be internally elaborated by following some functional requirements.

Using the mask’s properties that have been introduced via its defined rules, relations to other components, and effect on the architectural layout, we introduce some basic functional requirements:
\begin{enumerate}[label=\textbf{F\arabic*}, leftmargin=*]
\item \label{f1} The mask must interface with the system via mediators. The mask connects to just one mediator, but it should in general be able to connect to and communicate with any system mediator interchangeably. A connection with a wrapper is feasible but, it is inadvisable and thus not of primary concern.
\item \label{f2} The mask must provide a user access interface. The user access interface is the point of user system access. This interface can take any implementational form, provided that the chosen form has presentational abilities for data storage concerns. This interface is interchangeable and does not have effect on the general way in which data source component translations take place.
\item \label{f3} The mask must translate schemas from the system format to the user access (masked) format. The mask ascertains the system schema provided by its connected mediator and adapts the schema to a defined mask format.
\item \label{f4} The mask must translate queries from the user access (masked) format to the system format. The queries are given by the user through the user access interface in a masked format and are translated to the system format. To determine mask-to-system element mappings, the query translation can use the results of schema translations.
\item \label{f5} The mask must translate results from the system format to the user access (masked) format. The results received through the system must be adapted to the defined mask format. To determine certain metadata aspects (ex. the naming of attributes) of data results, can use the results of schema translations.
\end{enumerate}

The requirement \ref{f1} follows from \ref{rma1}, \ref{f2} from \ref{rma2}, and the requirements \ref{f3}, \ref{f4} and \ref{f5} from \ref{rma3}.

\begin{figure}[h]
\centering
\includegraphics[width=0.8\textwidth]{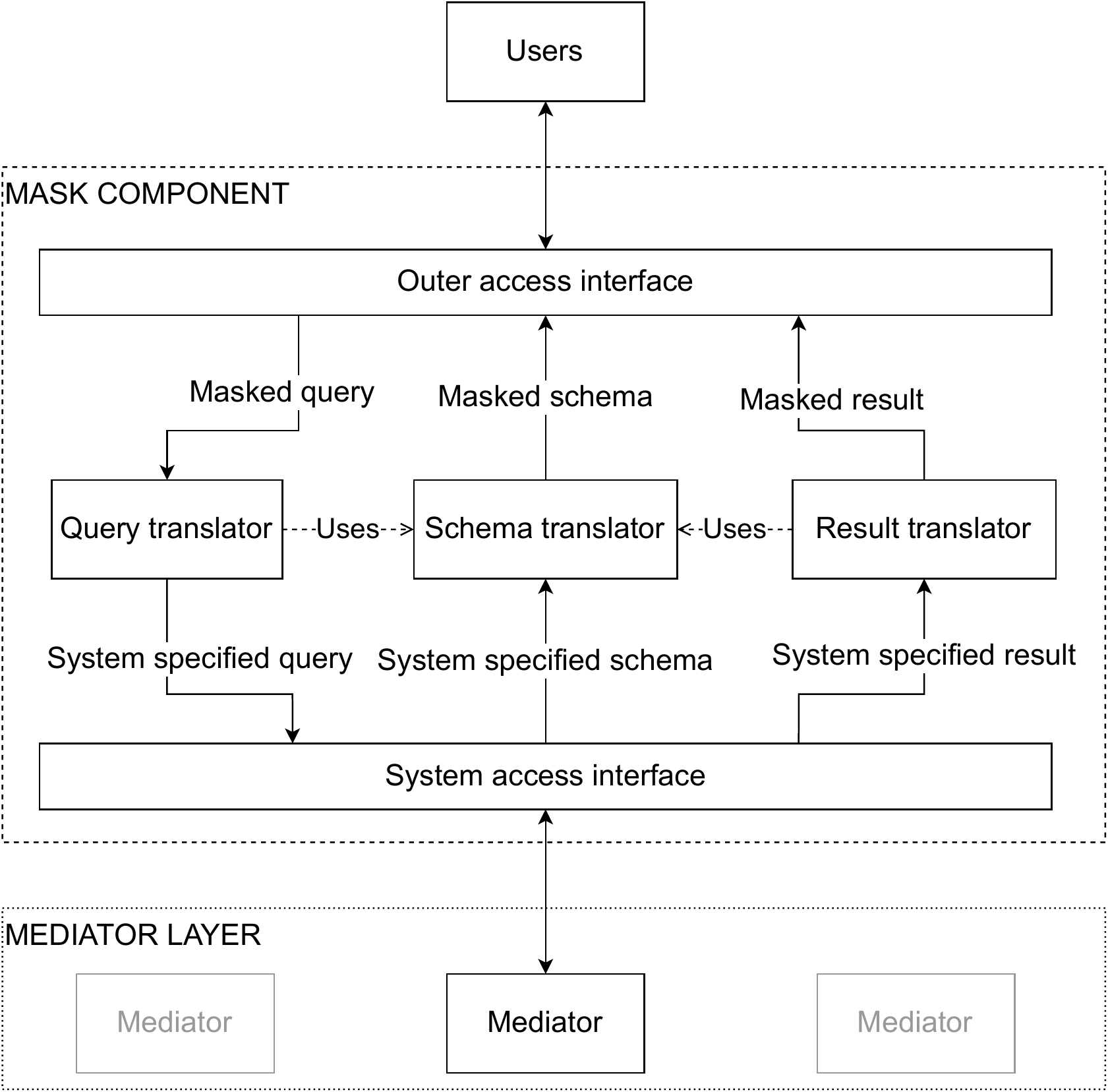}
\caption{A conceptual model of the mask’s functional components}\label{fig10}
\end{figure}

Following these functional requirements, a conceptual depiction of the mask’s inner components is devised. This is displayed in Fig.~\ref{fig10} as a conceptual model of functional components. These components present a generalized idea of what kind of functionalities should a mask have and what should their relationships be in terms of data exchange and dependency. These components do not present real-world components, but rather a possible grouping of some real-world components providing a functionality.

This sketch allows the mask’s functionalities to be put into context. The schema, query, and result translators are recognized as components with the task of translating data source components. The central role in translation is given to the schema translator as queries and query results are translated by using schemas generated by the schema translator. The system access interface is used to connect to the system via a mediator in the layer below. The outer access interface is a generic component, able to accommodate an adequate form of an access interface.

A noticeable trademark of this model is that there is a focus on the flow of data and its conversion by the components. A masked query is translated and sent (down) into the system. Reciprocally, the result of such a query is translated into a masked format to be sent (up) to the user. Similar is the case of schema translation, the system specified schema is translated into a masked schema for presentation to the user.

\begin{figure}[h]
\centering
\includegraphics[width=\textwidth]{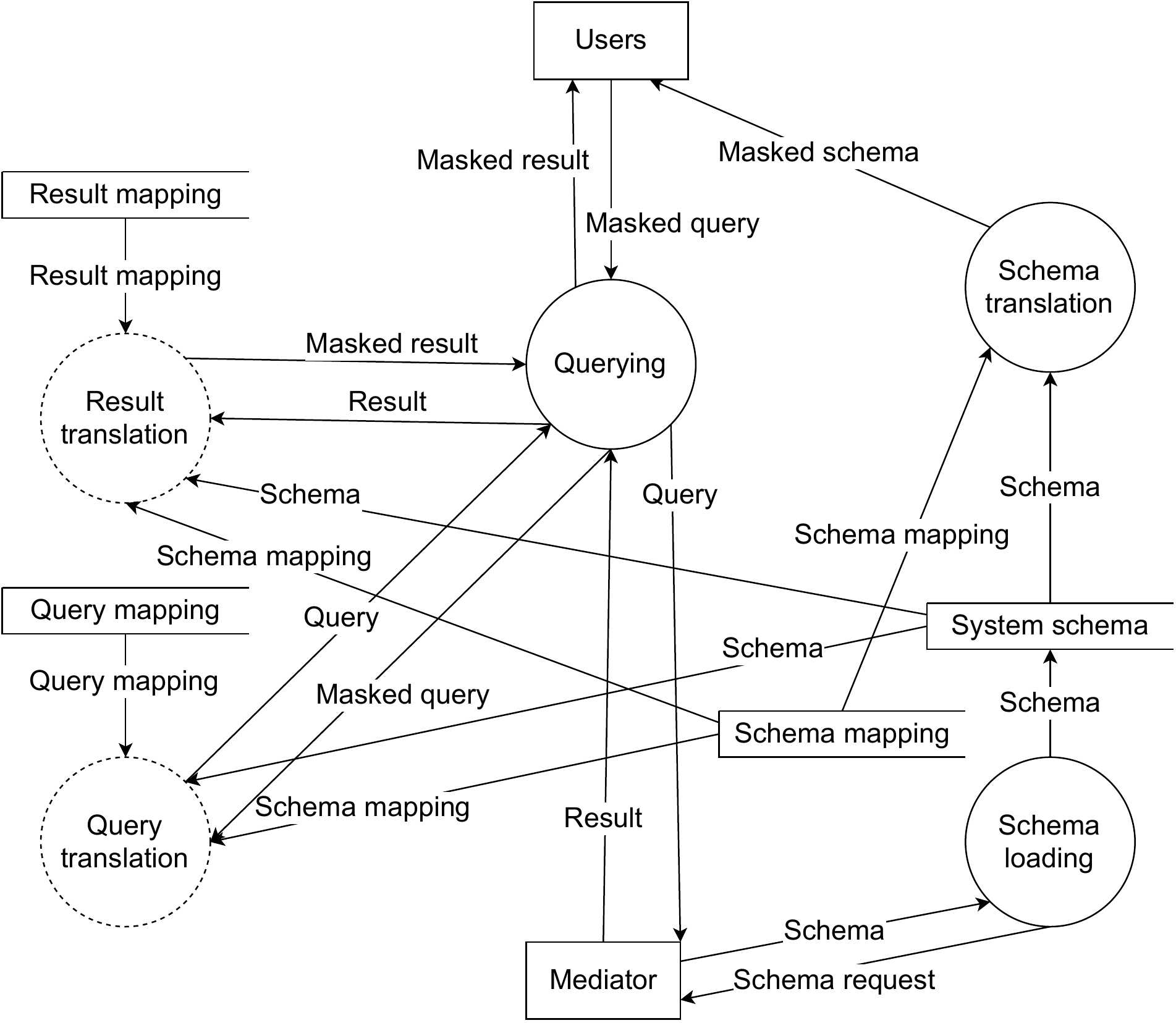}
\caption{Dataflow diagram following the mask’s functionalities}\label{fig11}
\end{figure}

Such data transformations can only be achieved through processes, so in a general sense it is more sensical to discuss the mask in terms of processes and the data that flows between them. To achieve a more detailed elaboration of the mask, building upon the model from Fig.~\ref{fig10}, a data flow diagram is constructed as displayed in in Fig.~\ref{fig11}.

On the right-hand side of Fig.~\ref{fig11}, the schema translator of Fig.~\ref{fig10} is decomposed as two processes: schema loading and schema translation. The schema loading process is concerned with the acquisition of a system schema from a connected mediator. This schema is also stored for other usage, besides schema translation, but it should be reacquired frequently to maintain an up-to-date schema. For this reason, schema loading is considered a separate and independent process. The schema translation process uses the currently acquired system schema and schema mapping rules from a separate storage to create a masked schema. This masked schema is presented to the user.

The querying process in Fig.~\ref{fig11} is a complex process that concerns itself with querying over a mediator. It is closely tied with processes of query translation and result translation. These processes are effectively subprocesses of querying but have been extracted due to their importance and correlation with components in Fig.~\ref{fig10}. 

The query translation process translates a masked query into a system formatted query. It requires data about the current system schema and schema mapping rules to determine the way in which they are reflected onto the current query. This must be considered as the schema translation might change resources’ names or change their schematic, so it becomes important to reverse those translations when constructing a system query. The query translation process also requires query mapping rules. As a general example, and for the moment setting a simplified generic model for a query – these rules might explain how a projection or selection in a query are to be translated.

The result translation process also requires data about the current system schema and schema mapping rules, as it is also concerned with translating a small view-like portion of the schema with the addition of holding result data that can also go through some masking transformations. Just like in other translational cases, result data translation also requires some mapping rules for data results.

There is also a very interesting feature of the diagram in Fig.~\ref{fig11} regarding all the mapping (rules) data storages. The result mapping, schema mapping, and result mapping data storages do not have any data inflows. These mappings are, in the context of this diagram, then clearly provided by some other undefined source. In fact, these mappings can only be provided by the developers of a certain mask component. These mappings are the exact point at which the system can no longer be designed as generic or abstract, and some concrete implementation or empirical data describing the masking of the system is required.

\begin{figure}[h]
\centering
\includegraphics[width=\textwidth]{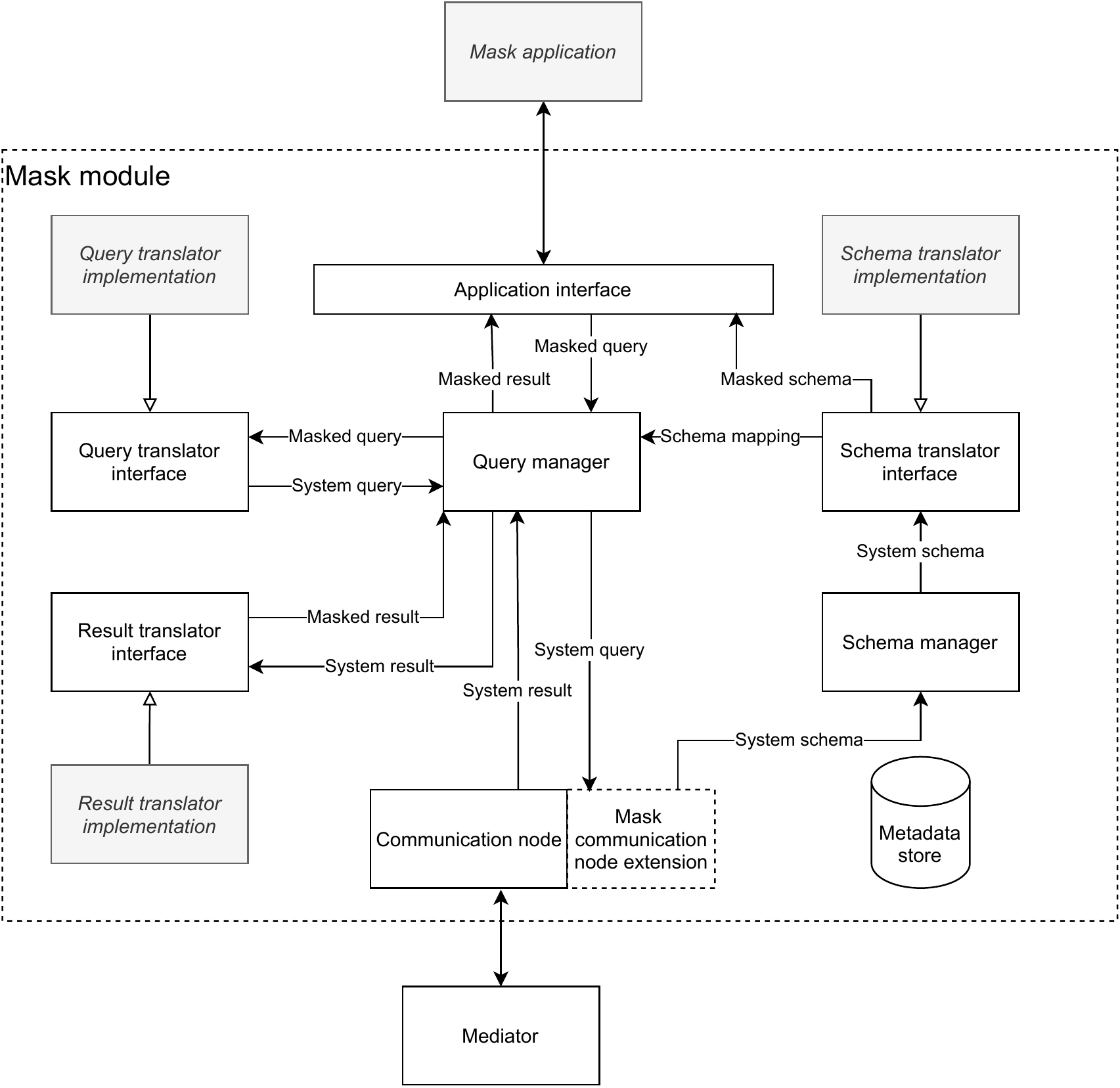}
\caption{Conceptual component design of the mask component}\label{fig12}
\end{figure}

Considering the mentioned findings, a conceptual component design for a mask is proposed and shown in the diagram of Fig.~\ref{fig12}. The goal is to also think of the mask’s component design without reducing generality to avoid prescribing any concrete programming paradigms, languages, or specific design patterns.

Following the process inference of the data flow diagram of Fig.~\ref{fig11}, in Fig.~\ref{fig12} the schema manager and schema translator (interface) are recognized as components – analogous to the schema loading and schema translation processes, respectively. The schema manager observes the system schema and updates the stored system schema appropriately.

The inference of a general querying process in the data flow diagram leads to the recognition of the query manager component. This component manages the underlying translations and query execution sequencing. In general, its purpose is to produce a masked result for a masked query. This is achieved through the processes of query and data result translation, which themselves in turn lead to the recognition of the query translator (interface) and data result translator (interface) components.

Regarding the mask’s communication abilities with the rest of the system – the mask sends system format queries and results, just like mediators and wrappers. For this reason, the mask can use a standardized communication node (module) used in all other component types. Due to the communication restrictions of the mask (allowing for a single mediator connection), an extension of a basic communication node should also be implemented. This also allows the addition of new message types if such a requirement should arise later.

To store all the data inferred for storage in Fig.~\ref{fig11}, a metadata store should also be introduced to the component. Such a store would be used by all components that require at least some schematic information or technical information. In Fig.~\ref{fig12} this store is displayed, but the connections to other components are omitted for the sake of clarity.

On the other hand, the mappings (marked as data storages in Fig.~\ref{fig11}) are not stored in the metadata database. They are presented by the component implementations (marked gray) in Fig.~\ref{fig12}. For the mask to remain as generic as possible, such mapping rules are explicitly described by the schema, query, and result translator implementations. The aforementioned interfaces are used to keep a level of abstraction towards the other inner components. The implementations are case specific and created by the mask’s developers according to their respective interfaces. This allows the development of a mask kind to be done without the need for extensive coding, as only the missing implementation pieces need to be filled in. This, by consequence, not only simplifies the development process, but also decreases the time required to develop a certain kind mask component.

It is important to note that the term “interface” is used in its broadest form here; not excluding the development of the mask component in a non-object-oriented paradigm language (non OOP). Along with these interfaces being implemented as standard OOP interfaces, they can also be implemented as high-order functions in a functional paradigm or as separate implementations of function prototypes defined in library headers (to be linked before compilation) in a structural-procedural paradigm. This is one of the beneficial results of generic and abstract reasoning about mask components.

The inner components that have been elaborated up to this point are part of a mask module, or rather a library. This is best understood from the point of another component marked gray in Fig.~\ref{fig12} – the mask application. The mask application is the execution entry point of the mask component. In the continuation of previous possible use-case examples, this component could be a web API or a TCP server listening for JDBC. Whichever the exemplified case, it would use the mask module as a library to connect to the integration system. The interfacing of the mask application and the module is achieved through the mask application interface, that provides a universal interface for data storage. In essence the mask application interface provides: 
\begin{itemize}
    \item the acquisition of a mask schema
    \item querying via a masked query
    \item receiving masked results
\end{itemize}

A well-designed mask module allows developers to treat it as a simple native data provider without the need for additional transformations. Of course, the achievement of such a property is also dependent on the developer’s ability to provide schema, query, result translator implementations fitting well with the implemented mask application.

If such design generalizations were not considered, the development of each kind of mask component would create a lot of excess repeated work and increase the overhead workload as all aspects of a mask would need to be reimplemented and retested. Such development would also have an impact on the management of multiple mask kind codebases, as none would conform to any design standard.

Keeping in form with the proposed design, the development of a mask type component is narrowed down to the implementation of just four components:
\begin{itemize}
    \item schema translator implementation
    \item query translator implementation
    \item result translator implementation
    \item mask application
\end{itemize}

This obviously reduces the workload and time required to implement a mask component, removing the need for reimplementation of core components. The development along the proposed design allows logical layering of the mask component in the segment of the mask application, as the mask module can be treated as a provider or service. Such standardization allows the mask components to be potentially built, tested, and maintained by a community of developers in a form of an open-source software initiative.

\subsection{Quantitative analysis on scenarios}

To prove that the MMW architecture simplifies a MW-based data source integration system’s maintenance and change management, a leveled quantitative analysis to compare the MMW and MW alternatives is needed. The following analysis is based on an evolution-cost quantitative analysis for measuring software flexibility described by \citectx{eden_measuring_2006}.

\citectx{eden_measuring_2006} proposed that a software’s flexibility can be measured and compared to other designs by approximating the cost of implementing anticipated changes – shifts. The cost of shifts is defined as the quantity of software units that need to be changed, added, or removed. These software units are called modules in a general sense but are exemplified with classes and methods in the paper.

\begin{figure}[h]
\centering
\includegraphics[width=\textwidth]{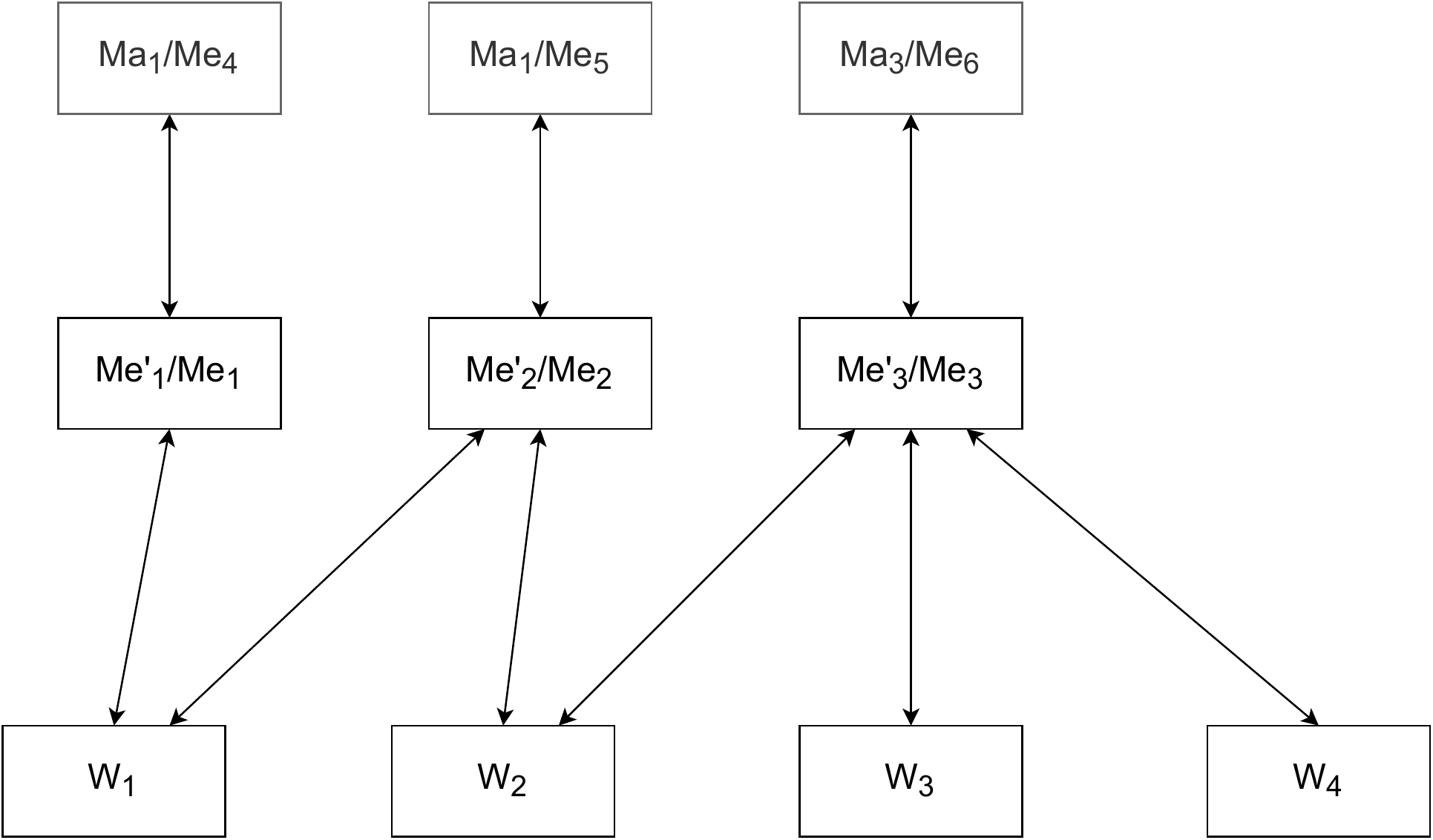}
\caption{Architecture used in the analysis}\label{fig13}
\end{figure}

To adjust this analysis for the level of architecture design in this paper, the modules are viewed as architectural components. The analysis compares an isomorphic example (shown in Fig.~\ref{fig13}) of a MW architecture with one mediator layer (1LMW), a MW architecture with two mediator layers (2LMW), and a MMW architecture. In the cases of 1LMW and 2LMW, mediators are considered to have functionalities of both mediation and representation.

The analysis is done over four scenarios: adding a new representation type, adding a new representation, adding a new mediator, and adding a new wrapper to a mediation. The symbolic nomenclature for this analysis is defined as follows:

\begin{quote}
    For a set of components $S_{comp}$ of possible types $S_{comptypes}=\{Ma, Me, Me', W\}$ representing a mask, a mediator with representational functionality, a mediator without representational functionality, and a wrapper respectively, and a set of possible actions over those components $S_{act}=\{impl,depl\}$ representing implementation and deployment respectively, $C_{X}^Y$ us the cost of performing an action $Y \in S_{act}$ over component $X \in S_{comp}$, with the addition of $C_{Conn}^{set}$ signifying the cost of setting up a connection between a pair of components $\{(c_1,c_2) \mid c_1,c_2 \in S_{comp}\}$
\end{quote}

Since a mediator with representational functionality is more complicated to implement than a mediator without representational functionality, the cost of implementing the former is greater than the latter:

\begin{equation}\label{eq1}
C_{Me}^{impl} > C_{Me'}^{impl}    
\end{equation}

Due to a greater number of functionalities that need to be supported by the surrounding system to which the component is being deployed, the deployment of a mediator with representational functionality is also more costly than that of a mediator without representational functionalities. This is because their deployment includes the tasks of setting up system resource access permissions, component settings, and firewall rules; all of which are either increased in quantity or complexity in the case of a representational mediator. Therefore, we conclude the following expression: 

\begin{equation}\label{eq2}
C_{Me}^{depl} > C_{Me'}^{depl}    
\end{equation}

As in the former statement, for the same reasons the deployment of a mask component is considered less costly than a mediator with representational functionalities. In addition, the mediator has a communication node intended for access to multiple sources. This is considered bloat, as the representational components connect to only one component in the lower layer. The mask, on the other hand, has a communication node inherently allowing just one connection to the lower layer (as per \ref{rma2}), making the connection configuration simpler. Therefore, we conclude the following expression:

\begin{equation}\label{eq3}
C_{Me}^{depl} > C_{Ma}^{depl}    
\end{equation}

\subsubsection{Scenario 1: Adding a new representation type}

\begin{figure}[h]
\centering
\includegraphics[width=\textwidth]{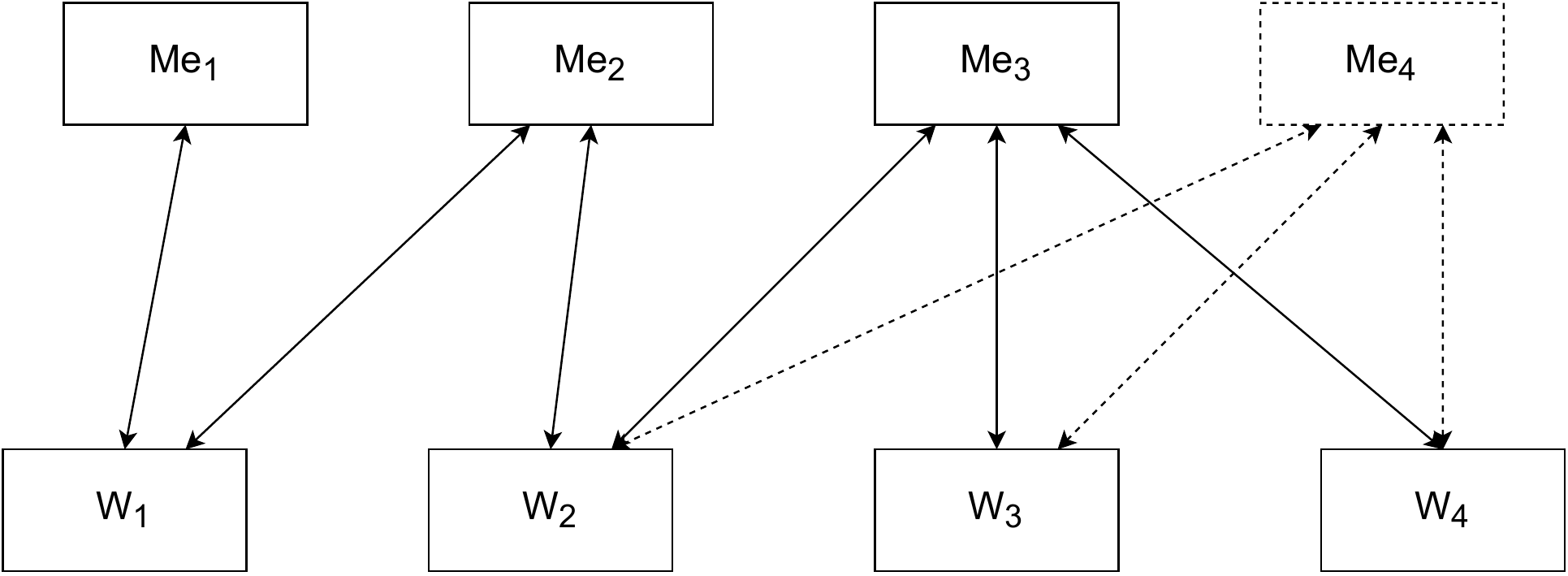}
\caption{Scenario 1 outcome on a one-layer mediator MW architecture}\label{fig14}
\end{figure}

In this scenario, a requirement of for a new representation type on top of combined schemas of wrappers $W_2$, $W_3$, and $W_4$ is added.
Since a new type of representation is required, in a 1LMW an entirely new mediator must be implemented. This new mediator also must be deployed and connected to wrappers $W_2$, $W_3$, and $W_4$. The outcome of the shift is displayed in Fig.~\ref{fig14} with the addition of $Me_4$ and its connections to the required wrappers.

Thus, the cost of the shift is:

\[C_{1LMW}^1=C_{Me}^{impl} + C_{Me}^{depl} + 3 \times C_{Conn}^{set}\]

In a general case, with the number of connected wrappers being $N$, the cost is:

\[C_{1LMW}^1=C_{Me}^{impl} + C_{Me}^{depl} + N \times C_{Conn}^{set}\]

It can be noticed that this architecture forms redundant connections between wrappers and mediators, adding to the shift cost.

\begin{figure}[h]
\centering
\includegraphics[width=\textwidth]{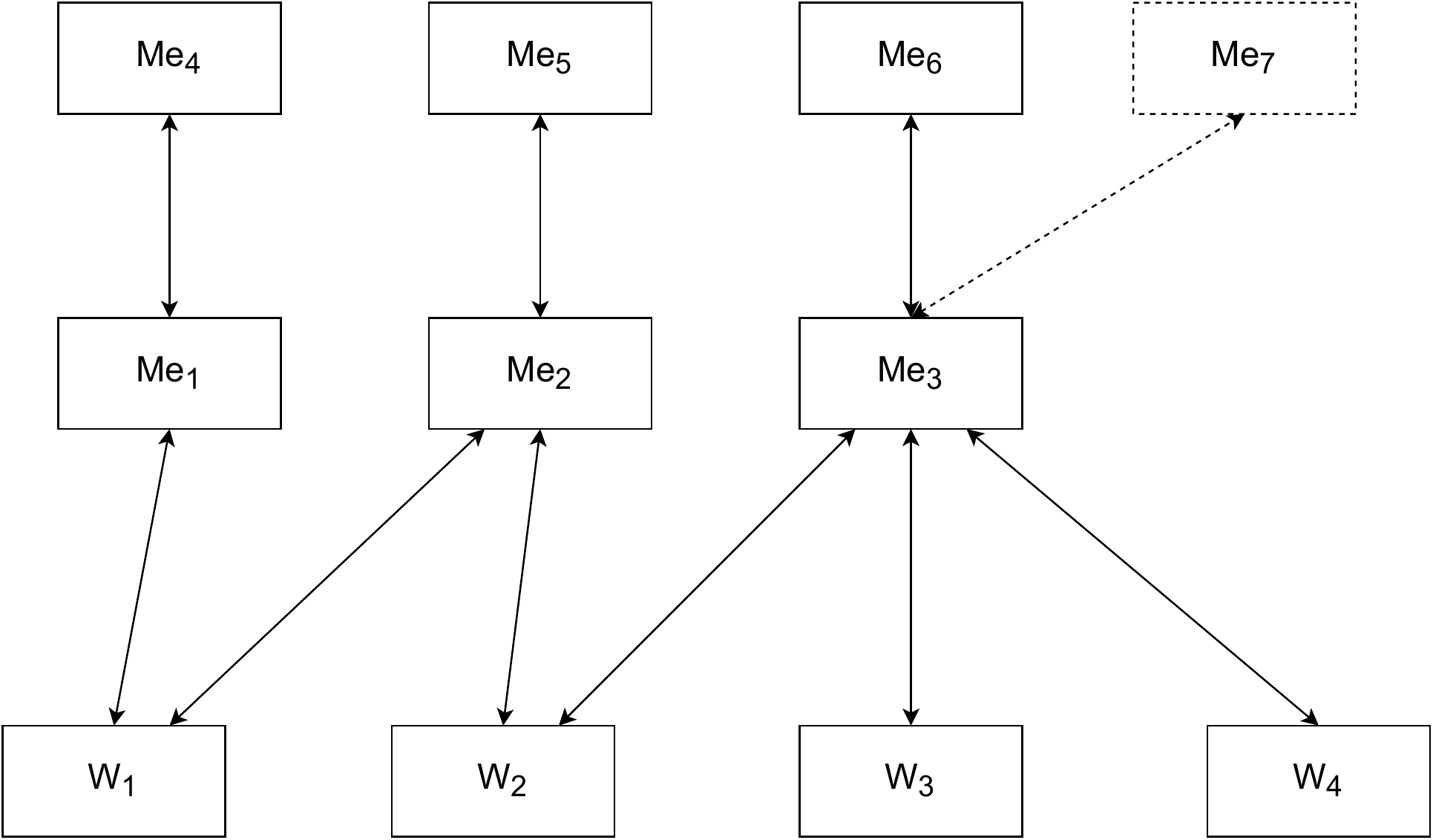}
\caption{Scenario 1 outcome on a two-layer mediator MW architecture}\label{fig15}
\end{figure}

Again, the case of a 2LMW, a new mediator must be implemented and deployed ($Me_7$). In this case mediator is stacked on top of a mediator on the lower layer of mediators ($Me_3$). Hence, the mediator $Me_3$ is reused for combining wrappers $W_2$, $W_3$, and $W_4$ and only one connection is set up. The outcome of the shift is displayed in Fig.~\ref{fig15}. 

The cost for this shift is:

\[C_{2LMW}^1=C_{Me}^{impl} + C_{Me}^{depl} + C_{Conn}^{set},\] which remains true for any general case.

\begin{figure}[h]
\centering
\includegraphics[width=\textwidth]{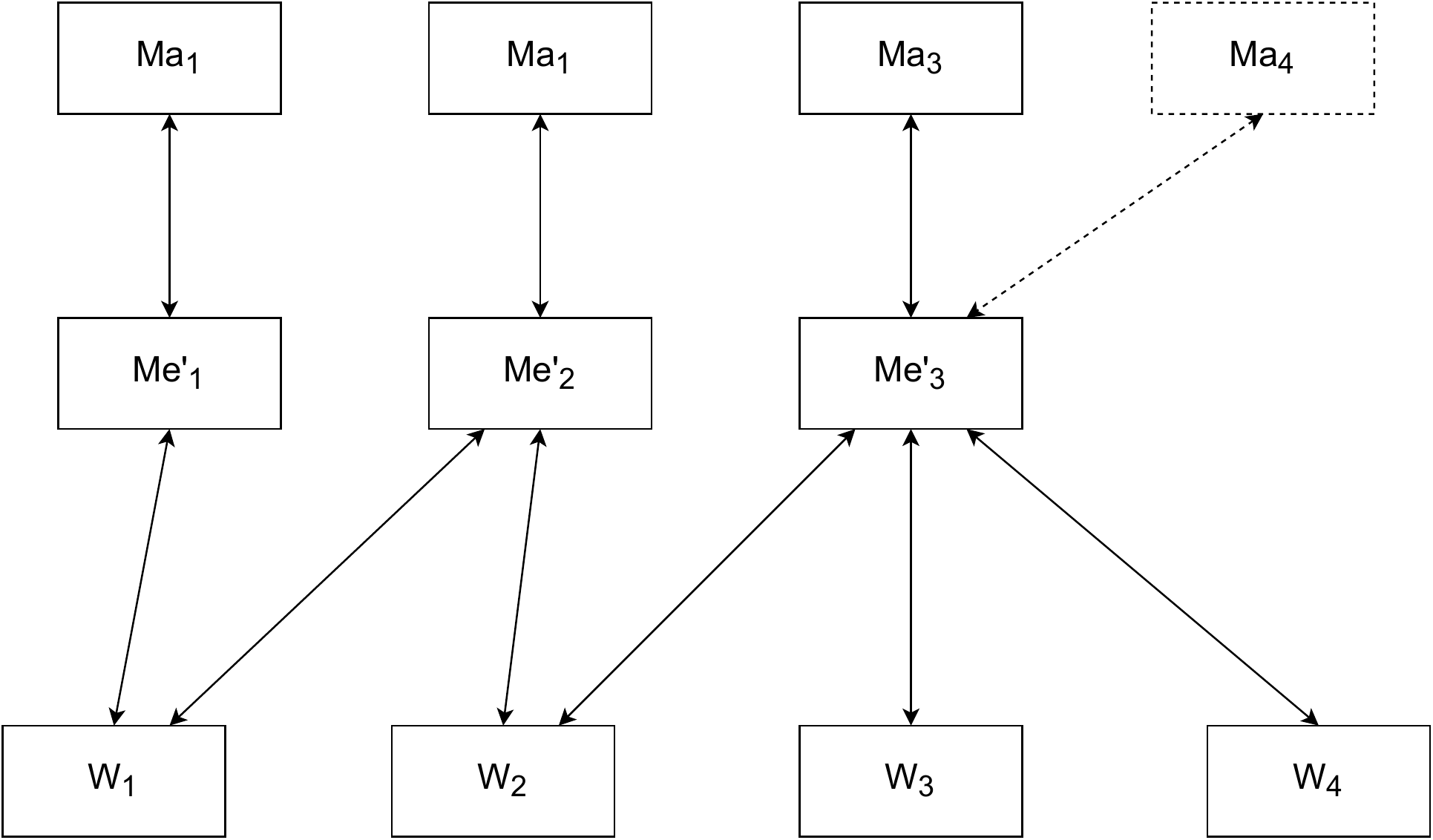}
\caption{Scenario 1 outcome on a MMW architecture}\label{fig16}
\end{figure}

In the case of a MMW architecture. To create a new type of representation, a new mask ($Ma_4$) is required to be implemented and deployed. Only one connection setup is required as the new mask only connects to one mediator in the mediator layer ($Me’_3$). It is important to note that the mediators in this architecture don’t serve a representational purpose, so they don’t have representational functionality. The outcome of the shift is displayed in Fig.~\ref{fig16}.

The shift cost in this and general cases is:

\[C_{MMW}^1=C_{Ma}^{impl} + C_{Ma}^{depl} + C_{Conn}^{set}\]

\subsubsection{Scenario 2: Adding a new representation}

In this scenario a requirement for a new representation on top of combined schemas of wrappers $W_1$ and $W_2$ is added. The representational component is already implemented, so none of the cases will have an implementation cost, just the deployment cost and the cost of connection setup.

In a 1LMW the new mediator is deployed and two connections to the two wrappers $W_1$ and $W_2$ set up. The shift cost is:

\[C_{1LMW}^2=C_{Me}^{depl} + 2 \times C_{Conn}^{set}\]

In a general case the shift cost is determined by the number of required redundant connections to wrappers N:

\[C_{1LMW}^2=C_{Me}^{depl} + N \times C_{Conn}^{set}\]

In a 2LMW a new mediator is deployed and one connection to its underlying mediator ($Me_2$) set up. The shift cost is:

\[C_{2LMW}^2=C_{Me}^{depl} + C_{Conn}^{set}\]

In a MMW architecture a new mask is deployed and one connection to its underlying mediator ($Me’_2$) set up. The shift cost is:

\[C_{MMW}^2=C_{Ma}^{depl} + C_{Conn}^{set}\]

\subsubsection{Scenario 3: Adding a new mediator}

In this scenario a requirement for a new mediator over wrappers $W_2$ and $W_3$ is added. It will be assumed that this type of mediator already exists, so there is no cost of implementation.

In a 1LMW a mediator is deployed and connected to the two wrappers.  the shift cost is:

\[C_{1LMW}^3=C_{Me}^{depl} + 2 \times C_{Conn}^{set}\]

Again, for a general case where $N$ is the number of connected wrappers, the shift cost is:

\[C_{1LMW}^3=C_{Me}^{depl} + N \times C_{Conn}^{set}\]

In a 2LMW a mediator must be deployed to the lower mediator layer to combine the wrappers and a mediator in the upper mediator layer to provide the representation. The set-up connections also must be considered, as two connections are set up towards the wrappers and a single connection between the mediators. The shift cost is:

\[C_{2LMW}^3= 2 \times C_{Me}^{depl} + 3 \times C_{Conn}^{set}\]

In a general case, where $N$ is the number of connected wrappers, the shift cost is:

\[C_{2LMW}^3= 2 \times C_{Me}^{depl} + (N+1) \times C_{Conn}^{set}\]

In a MMW architecture, together with deploying a mediator, a mask must be provided. The mask type is considered as already implemented (analogous to the cases of MW architectures), so it has to only be deployed. Two connections are set up towards the wrappers and a single connection between the mediator and mask. The shift cost is:

\[C_{MMW}^3=C_{Me'}^{depl} + C_{Ma}^{depl} + 3 \times C_{Conn}^{set}\]

In a general case, where $N$ is the number of connected wrappers, the shift cost is:

\[C_{MMW}^3=C_{Me'}^{depl} + C_{Ma}^{depl} + (N+1) \times C_{Conn}^{set}\]

\subsubsection{Scenario 4: Adding a new wrapper to an existing mediation}

Additionally, to demonstrate that these architectures are sound (the MMW architecture first and foremost), a scenario of adding a new wrapper can be analyzed. The appending of wrappers to an existing mediator doesn’t impact the rest of the components, as the wrapper is deployed and a single connection to the required mediator set up. Thus, the shift cost for all architectures is:

\[C_{1LMW}^{4}=C_{2LMW}^{4}C_{MMW}^{4}=C_{W}^{depl}+C_{Conn}^{set}\]

\subsubsection{Analysis of the shift costs}

With the shift costs evaluated, a more concise comparison of architectures can be made. Table~\ref{tab1} displays all the shift costs for each scenario and architecture.

\begin{table}[]
\caption{Shift costs for all scenarios and architectures}\label{tab1}
\centering\resizebox{\textwidth}{!}{ 
\begin{tabular}{|l|l|l|l|}
\hline
\textbf{Sc.} & \textbf{1LMW} & \textbf{2LMW} & \textbf{MMW} \\ \hline
1            & $C_{Me}^{impl} + C_{Me}^{depl} + N \times C_{Conn}^{set}$    & $C_{Me}^{impl} + C_{Me}^{depl} + C_{Conn}^{set}$    & $C_{Ma}^{impl} + C_{Ma}^{depl} + C_{Conn}^{set}$   \\ \hline
2            & $C_{Me}^{depl} + N \times C_{Conn}^{set}$    & $C_{Me}^{depl} + C_{Conn}^{set}$    & $C_{Ma}^{depl} + C_{Conn}^{set}$   \\ \hline
3            & $C_{Me}^{depl} + N \times C_{Conn}^{set}$    & $2 \times C_{Me}^{depl} + (N+1) \times C_{Conn}^{set}$    & $C_{Me'}^{depl} + C_{Ma}^{depl} + (N+1) \times C_{Conn}^{set}$   \\ \hline
4            & $C_{W}^{depl}+C_{Conn}^{set}$    & $C_{W}^{depl}+C_{Conn}^{set}$    & $C_{W}^{depl}+C_{Conn}^{set}$   \\ \hline
\end{tabular}
}
\end{table}

The first scenario demonstrates that in the MMW architecture the addition of a new type of representation is only dependent on the implementation and deployment of a mask component. The other two architectures depend on mediator components. The 1LMW’s shift cost noticeably depends on the number of connected wrappers – to emphasize: for adding a representation type. The 2LMW and MMW architectures are not at such a disadvantage, their difference being the type of component added to the system. Since a mask is less costly to implement and deploy than a mediator, the overall shift cost in scenario 1 is lowest in the MMW case.

The second scenario also shows that the 1LMW shift cost is dependent on the number of wrappers. The cases of 2LMW and MMW are again analogous, but a mask is less costly to deploy than a representational mediator. This makes the shift cost of the MMW case the lowest again. As it was discussed earlier in the text, using a mediator just for representation, without using its mediation functionalities, is akin to killing a fly with a cannonball. 

The third scenario shows the shift cost overhead the 2LMW and MMW have as opposed to the 1LMW when setting up mediation. There is an obvious trade-off in these architectures between the shift cost of adding mediation or representations. To maintain a less costly (and qualitatively simpler) representation addition, the overhead cost of adding mediation is increased. This overhead can be quantified for 2LMW as:

\[C_{2LMW}^{overhead}=C_{2LMW}^{3}-C_{1LMW}^{3}=C^{depl}_{Me}+C_{Conn}^{Set},\] and for the MMW:

\[C_{MMW}^{overhead}=C_{MMW}^{3}-C_{1LMW}^{3}=(C_{Me'}^{depl}-C_{Me}^{depl})+C_{Ma}^{depl}+C_{Conn}^{set}\]

Considering that the MMW mask and mediator are less costly to deploy than the 2LMW mediator, the overhead cost is reduced in favor of the MMW.

The fourth scenario shows that the addition of a new wrapper to the MMW system has no effect on the rest of the system hierarchy, as it is also expected of the other MW architectures. The MMW finds itself in no detrimental opposition to the other architectures.

\subsection{Hypothetical implementational example}

As referenced before, \citectx{atzeni_uniform_2014} presented the SOS (Save Our Systems) as a system for uniform operations over non-relational stores. The representational format of this system was a web API with URI-like resource identification, serving result data as JSON objects. They have also defined generic wrappers (in their case as modules) which are implemented per data store kind. The wrappers in their case also concern themselves with representing the underlying systems. Although not a data source integration system itself, the SOS system could keep its raison d'être and be conveniently extendable to the likes of a data source integration system if it were reimagined following the MMW architecture. Additionally, using different mask kinds, the representational form of these data storages could be expanded on.

\begin{figure}[h]
\centering
\includegraphics[width=0.75\textwidth]{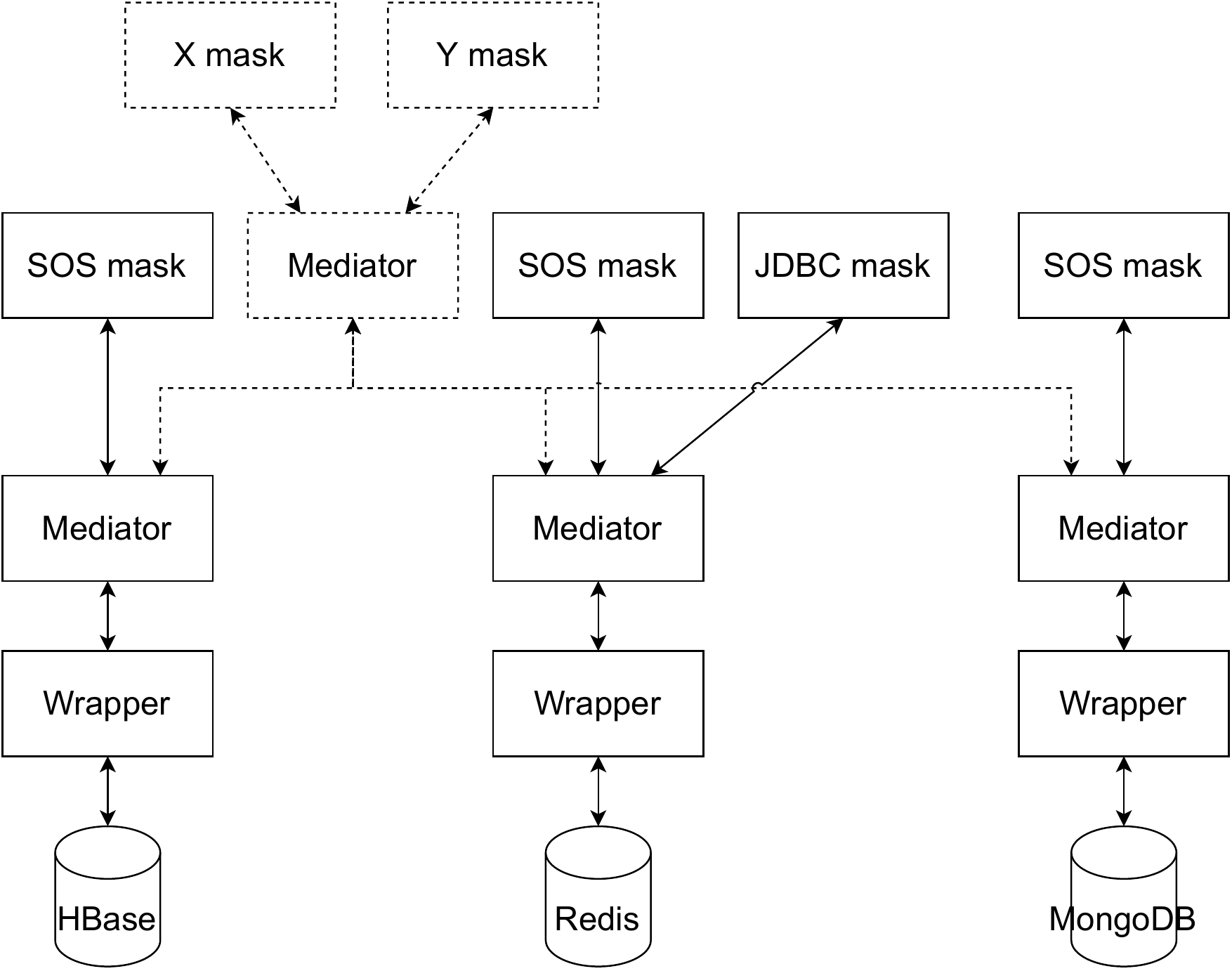}
\caption{The SOS system implemented in the MMW architecture}\label{fig17}
\end{figure}

A hypothetical example of such a system use case is presented in Fig.~\ref{fig17}, where the revised SOS system provides uniform access over a single HBase, Redis, and MongoDB database. At the top of the component hierarchy each database must be converted to the SOS interface for application access, as is presented by \citectx{atzeni_uniform_2014}. This interface can be presented via mask components of the SOS mask kind.

Starting from the bottom of the Fig.~\ref{fig17} hierarchy, a wrapper is connected to each data source. Since it is not advisable to connect masks directly to wrappers and following RMa2 stating that masks should only connect to mediators – a single mediator is connected to each wrapper. A SOS mask is connected to each of these mediators, thus encompassing the original SOS system’s functionalities. 
Also, these settings do not restrict the system to just using a SOS mask.  Additional mask kinds can be connected to these singular mediators to offer an alternative presentational form and keep in line with the SOS’s use-case of uniform system access. Such an example is given in Fig.~\ref{fig17} with a JDBC mask representing the Redis database.

The singular mediators are of a translational nature, but also enable the system hierarchy to be expanded. If, by example shown in Fig.~\ref{fig17}, a mediator connecting to each of the singular mediators is provided, then that part of the system becomes a data source integration system. Consequently, the integrating mediator allows connections coming from different masks. 

Not only does the MMW architecture allow the emulation of a system such as the SOS, but also expands on it. If it were not for the mask components taking on the responsibility of system representation, all mediators would have to be adapted (redesigned and reimplemented) to the SOS interface standard to work as both the SOS system and a data source integration system, which itself does not explicitly require adherence to the SOS interface standard.

\section{Conclusion}

The MW architecture has numerous advantages when it comes to implementing a data source integration system. Especially if the system is expected to connect a larger number of data sources. 
It was also demonstrated that, when certain rules are followed to keep the components well-formed, the maintenance of such systems also becomes easier. In the case of appending new data sources, they are each assigned a wrapper and the wrappers are connected to selected mediators. In the case of managing schemas, this management is modularized by the system architecture itself. Workloads on hardware can also be balanced, as components can be setup on different machines.

Although it has been mentioned that some authors did not use the MW architecture in terms of system components, most of them recognized these advantages and modeled their inner system modules to work in a similar fashion. This allowed them to at least use the advantageous modularization of tasks and responsibilities in a data source integration system. 

The architectural pattern’s idea has been agreed with up to a point. This point being the representation of data to users and the responsibilities of upper layer mediators. It has been found that there is a responsibility of representing data that usually has either been delegated to the upper layer of mediators or specialized user application. To keep this responsibility inside the system, the idea of a new architectural component has been proposed – the mask. As with mediators and wrappers, some basic rules for masks have been set to assure that they are well-formed and fit into a MW architecture.

The mask takes on the responsibility of representing data and creating an interface for users to connect to. It has also been reasoned that appending masks to a MW architecture technically creates a new architectural layer – the mask layer. 

By using the mask layer, a data source integration system increases its representational versatility and thus, its usefulness. Now a data source integration system can be viewed as a singular source of data, but through multiple masked sources. In addition, if masks adhere to a certain representational standard, then the data source integration system becomes open to different kinds of user applications.

The mask component was internally detailed as much as possible in an abstract and generic form to prevent any partiality to a specific programming language or paradigm. Reasoning about the mask’s hypothetical functional requirements has progressed to a simple functional component structure. This basic model was then enriched by recognizing processes that a mask should support, leading to a more detailed conceptual inner-component design. This process also allowed thinking about a more concrete placement of the mask component in a real-life implementation, which was concluded with the proposal that a mask should be implemented as a software module used by the mask component’s application code. The mask’s software module requires partial implementation, per mask-kind, for it to be used.

Because of the addition of a new architectural layer, it has been decided to name this a revision of the MW architecture and name it as a MMW architecture. To show that the MMW architecture has benefits over other MW architectures, a quantitative analysis over requirement shift costs was done on multiple scenarios. To showcase the possibilities of a real-world use of the MMW architecture and its benefits, a hypothetical example emulating an existing system for uniform data store access (SOS) was presented. This example showed the benefits garnered from the redistribution of responsibility among component types in the MMW architecture, specifically the responsibilities assigned to the mask. 

Further work is needed for additional detailing of the mask components and its interactions with other components (mainly mediators). It would also be interesting to investigate if there is a possibility of further generalization of the base implementation of a mask to achieve faster and standardized development of mask kinds. There is also a possible outlook for research on constructing mask generators, akin to wrapper generators implemented in referenced work. 
\bibliographystyle{ACM-Reference-Format}
\bibliography{MaskMediatorWrapper}

\end{document}